\documentclass[twocolumn]{aastex631}

\usepackage{subfigure}
\usepackage{soul}
\usepackage{etoolbox}
\robustify{\cite}
\robustify{\citep}
\usepackage[normalem]{ulem}

\usepackage{graphicx} 
\usepackage{amsmath}
\usepackage{amssymb}
\newcommand{\angstrom}{\text{\normalfont\AA}}

\begin{document}

\title{The Great Escape of ionizing photons during Cosmic Morning}

\author[0000-0002-8768-9298]{Kanak Saha}\thanks{E-mail: kanak@iucaa.in}
\affiliation{Inter-University Centre for Astronomy and Astrophysics, Pune 411007, India.}

\author{Roland Bacon}
\affiliation{Univ Lyon, Univ Lyon1, Ens de Lyon, CNRS, Centre de Recherche Astrophysique de Lyon UMR5574, F-69230, Saint-Genis-Laval, France}

\author{Anne Verhamme}
\affiliation{Observatoire de Genève, Chemin Pegasi 51, CH-1290 Versoix, Switzerland}

\author{Soumil Maulick}
\affiliation{Inter-University Centre for Astronomy and Astrophysics, Pune 411007, India.}

\author{Akio K.\ Inoue}
\affiliation{Department of Physics \& Waseda Research Institute for Science and Engineering, Faculty of Science and Engineering, Waseda University, 3-4-1 Okubo, Shinjuku, Tokyo 169-8555, Japan}

\author{Souradeep Bhattacharya}
\affiliation{Centre for Astrophysics Research, Department of Physics, Astronomy and Mathematics, University of Hertfordshire, Hatfield AL10 9AB, UK}

\author{Marc Rafelski}
\affiliation{Space Telescope Science Institute, 3700 San Martin Drive, Baltimore, MD 21218, USA; Department of Physics and Astronomy, Johns Hopkins University, Baltimore, MD 21218, USA}

\author{Edmund Christian Herenz}
\affiliation{Inter-University Centre for Astronomy and Astrophysics, Pune 411007, India.}

\author{Rogier Windhorst}
\affiliation{School of Earth and Space Exploration, Arizona State University,Tempe, AZ 85287-6004, USA}

\author{Michael Rutkowski}
\affiliation{Minnesota State University-Mankato, Mankato, MN 56001, USA}

\author{Shyam N. Tandon}
\affiliation{Inter-University Centre for Astronomy and Astrophysics, Pune 411007, India.}


\begin{abstract}
The end of the Cosmic Dark Age marked the onset of reionization, driven by extreme-UV photons from the first galaxies. Direct detection of such photons has remained challenging due to strong intergalactic attenuation. Here, we report the first direct detection of ionizing photons at rest-frame wavelengths 350\AA, 392\AA, and 485\AA, using deep UV imaging from two independent space observatories: AstroSat and HST. These photons emerge from a stacked sample of spectroscopically confirmed Ly$\alpha$ emitters at $5.9<z<6.0$ in the Hubble Ultra Deep Field identified with VLT/MUSE and JWST/NIRSpec. These detections imply a higher transparency of the interstellar and intergalactic media to ionizing radiation than predicted by current models. The stacked spectrum, representative of faint galaxies with $M_{\mathrm{UV}}=-18.77\pm0.05$ ($\sim0.1L^{\ast}$), exhibits a hard slope ($-2.3\pm0.1$) and produces ionizing photons with $\text{log}_{10}\xi_{\text{ion}}^{true}$ = $25.86 \pm 0.02$~Hz\:erg$^{-1}$ and escape fraction $f_{esc}\simeq 0.8$. Individual measurements reveal very blue UV continua ($\beta \leq −3$ for three galaxies), young ages ($< 6$~Myr for four), and low metallicities ($Z\sim0.02$–$0.05\:Z_{\odot}$ for two), indicating the possible presence of very hot, massive stars capable of producing enough ionizing photons to drive cosmic reionization, thereby providing new constraints on its sources. Furthermore, detection of photons with energies $>24.6$~eV provide evidence that He{\sc i} reionization has begun by this epoch.
\end{abstract}

\keywords{galaxies: reionization; galaxies: high redshift; galaxies: ISM}

\section{Introduction}
\label{sec:intro}
Contemporary investigations into the escape of ionizing photons from galaxies rely primarily on indirect methods, focusing on the low-redshift analogs of reionization-era sources \citep{Madauetal1999}. These analog sources serve as empirical proxies for the high-redshift galaxies thought to have driven the cosmic reionization \citep{Naidu20,Atek24}, offering a tractable means of exploring the otherwise inaccessible conditions of the reionization-era universe \citep{Robertson22}. At present, a large sample of Lyman continuum (LyC) leakers at $z\sim 0.2 - 0.4$ have been identified using the Hubble Space Telescope’s Cosmic Origins Spectrograph, through the Low-redshift Lyman Continuum Survey (LzLCS) \citep{Fluryetal2022a} which includes analysis of earlier LyC leakers \citep{Izotov2016,Izotov16b,Izotov18,Izotov18b,Izotov21,Wang19}. Much of our current understanding of these low-z analogs has been gleaned from this survey, offering critical empirical insight into the conditions that enable LyC photon escape \citep{Flury22b}. Detections at intermediate redshifts ($z \sim 1 - 2$), however, remained rare \citep{Sianaetal2010, Rutkowskietal2016, Alavietal2020}. A notable exception is the discovery of an LyC emitter at $z\sim 1.42$ based on the deep imaging using AstroSat/UVIT \citep{Sahaetal2020}. Leveraging the deep UV imaging from the AstroSat UV Deep Field \citep{Sahaetal2024} and precise redshifts from MUSE \citep{Baconetal2023} and HST/Grism G141 \citep{Momvheva16,Simons23}, 17 LyC emitters \citep{Dhiwar24, Maulick25} have been subsequently identified at Cosmic Noon - a pivotal epoch of cosmic star-formation and stellar mass assembly \citep{ForsterSchreiber2020}. At relatively higher redshifts ($z \ge 2.4$), the Lyman limit shifts into HST’s WFC3/UVIS filters (F275W, F336W), enabling photometric identification of LyC sources \citep{Bianetal2017, RiveraThorsenetal2019, Fletcheretal2019, Yuanetal2021, Keruttetal2023, Smithetal2024}. Spectroscopic campaigns with Keck, VLT, and GTC have confirmed additional leakers at $z\sim 3 - 4$ \citep{Shapleyetal2016, Vanzellaetal2018, Steideletal2018, MarquesChavesetal2022, Liuetal2025}.

However, beyond $z \sim 4$, the transmission of the intergalactic medium (IGM) declines sharply, with the mean transmission (averaged over many sightlines) dropping to nearly zero by $z \sim 6$ \citep{Madau1995, Inoueetal2014}. As a result, the direct detection of ionizing photons near the Lyman limit ($911.8\ \AA$) from galaxies during the epoch of reionization remains extremely challenging and has largely eluded observational confirmation. Nevertheless, theoretical models suggest that IGM absorption is not uniform across all ionizing wavelengths shortward of the Lyman limit. In particular, the Lyman valley \citep{Moller90} — a region between $911.8\ \angstrom$ (13.6 eV) and $\sim 500\ \angstrom$ (24.79 eV) marks a trough in the transmitted UV flux, driven by cumulative absorption from intervening neutral hydrogen systems. Due to the decreasing photoionization cross-section ($\sigma_{HI} \propto \lambda^{3}$ for $\lambda < 911.8\ \angstrom$) of neutral hydrogen, higher energy photons are progressively less likely to be absorbed, even in the presence of abundant high-column-density systems. As a result, there is a modest but non-negligible probability of transmission below $\sim 500\ \angstrom$, even at $z \sim 6$. This opens the possibility that highly energetic ionizing photons ($\lambda < 500\AA$), capable of traversing the IGM, particularly along rare, low column density sightlines \citep{Madau1995, Inoueetal2014}, could be directly detected in deep UV imaging surveys that targets this spectral window. In this work, we present detection of such ionizing photons at $z\sim 5.94$ based on high-quality imaging and spectroscopic data from MUSE, HST, JWST and AstroSat. We further discuss the implications of our findings in the context of cosmic reionization.

The paper is organized as follows: section~\ref{sec:sample} describes the observations and data, with sample selection strategy detailed in section~\ref{sec:selection}. Section~\ref{sec:stack} describes the details of stacking procedure while section~\ref{sec:detection} presents the detection of ionizing photons, with offset and morphology of the stack being described in section~\ref{sec:stackmorph}. Robustness of the detection and various statistical tests are presented in section~\ref{sec:robust}. Details of the IGM modelling and the Lyman Valley transmission are described in section~\ref{sec:IGM}. The physical properties of invidual galaxies are discussed in section~\ref{sec:physical}. The escape fraction calculation is described in section~\ref{sec:fesc}. Section~\ref{sec:contamination} presents a detailed discussion on contamination calculation, while summary and implications are outlined in section~\ref{sec:summary}. Throughout this paper, we adopt the flat $\Lambda$CDM cosmology with $\rm H_{o}=70\: \text{kms}^{-1}Mpc^{-1},\ \Omega_{M} = 0.3 \ and \ \Omega_{vac} = 0.7$. All magnitudes quoted in the article are in the AB system \citep{Oke83}. 

\section{Observations, Data and Sample Selection}
\label{sec:sample}

\subsection{AstroSat/UVIT Observations}
\label{sec:astrosat}
The near-UV imaging data comes from the AstroSat UV Deep Field (AUDF) south in the N242W filter ($\sim 2000 - 3050 \AA$) observation \citep{Sahaetal2024}. The orbit-wise dataset was processed using the official L2 pipeline and the final science-ready image of the AUDF south has a total exposure time $t_{N242W}=62345$~sec corresponding to $\sim 63$~AstroSat orbits. The $3 \sigma$ depth of the final science-ready image is 27.73 AB mag and the full width at half maximum (FWHM) of the point spread function (PSF) is $\sim 1.2^{\prime\prime}$ \citep{Sahaetal2024}. The astrometry has an accuracy of 0.3$^{\prime\prime}$ RMS. Each pixel is 0.417$^{\prime\prime}$. 

\subsection{MUSE observations}
\label{sec:MUSE}
All galaxies in our stack sample are part of the MUSE Hubble Ultra Deep field surveys \citep{Baconetal2023} and are located within the MOSAIC dataset with an achieved depth of 10.8 hours and a spatial resolution of 0.64$^{\prime\prime}$ FWHM at 7000 \AA. Some of our galaxies lie within the MUSE eXtreme Deep Field (MXDF) that provides the highest sensitivity in the Lyman alpha detection (with $\sim 100$ hrs of integration). We use the latest data release DR2 \citep{Baconetal2023}, which benefits from an improved data reduction and analysis process. The measured accuracy of the MUSE astrometry is 0.08$^{\prime\prime}$ RMS with respect to the HST ACS images. 

\begin{figure*}
\begin{flushleft}
\rotatebox{0}{\includegraphics[width=1.\textwidth]{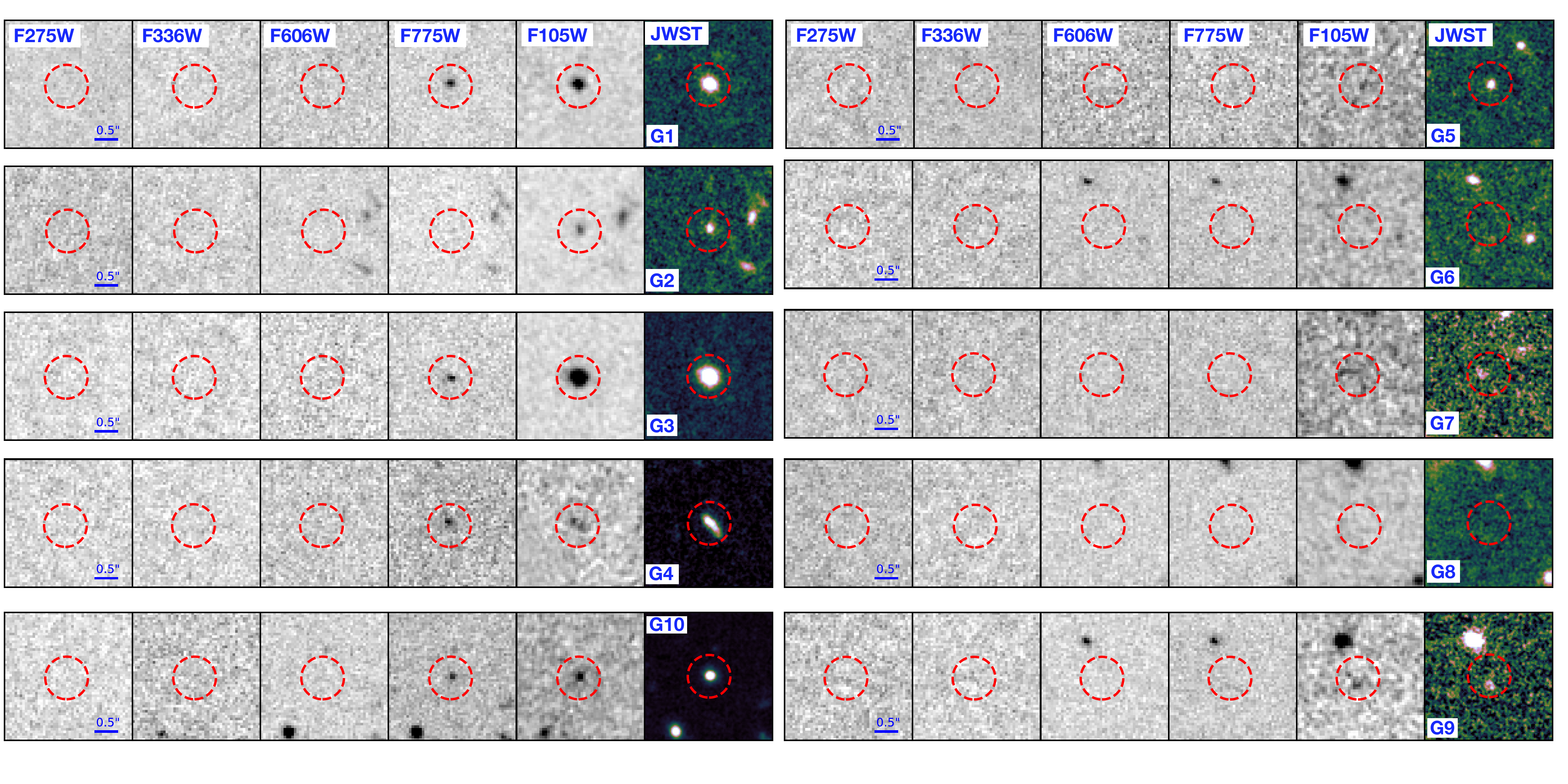}}
\caption{{\bf HST and JWST Postage Stamps:} Cutout (3$''$ x 3$''$) for all 10 galaxies (G1 - G10). G1, G2, G3, G4 and G10 are part of the HST Gold sample. G6 and G8 lack detectable stellar continuum either in the HST/Optical or JWST/NIRCam filters. Radii of the red dashed circular apertures are $0.5^{\prime \prime}$ and they are centered on the detected Ly$\alpha$ centroid of these galaxies from MUSE. The apparent offset between Ly$\alpha$ and stellar continuum (based on JWST detection image) in G7 and G9 are $0.2^{\prime \prime}$ in either case. All 10 galaxies together cover a spatial region of angular diameter $\sim 160"$.}
\label{fig:HSTstamp}
\end{flushleft}
\end{figure*}

\begin{figure}
\begin{flushleft}
\rotatebox{0}{\includegraphics[width=0.5\textwidth]{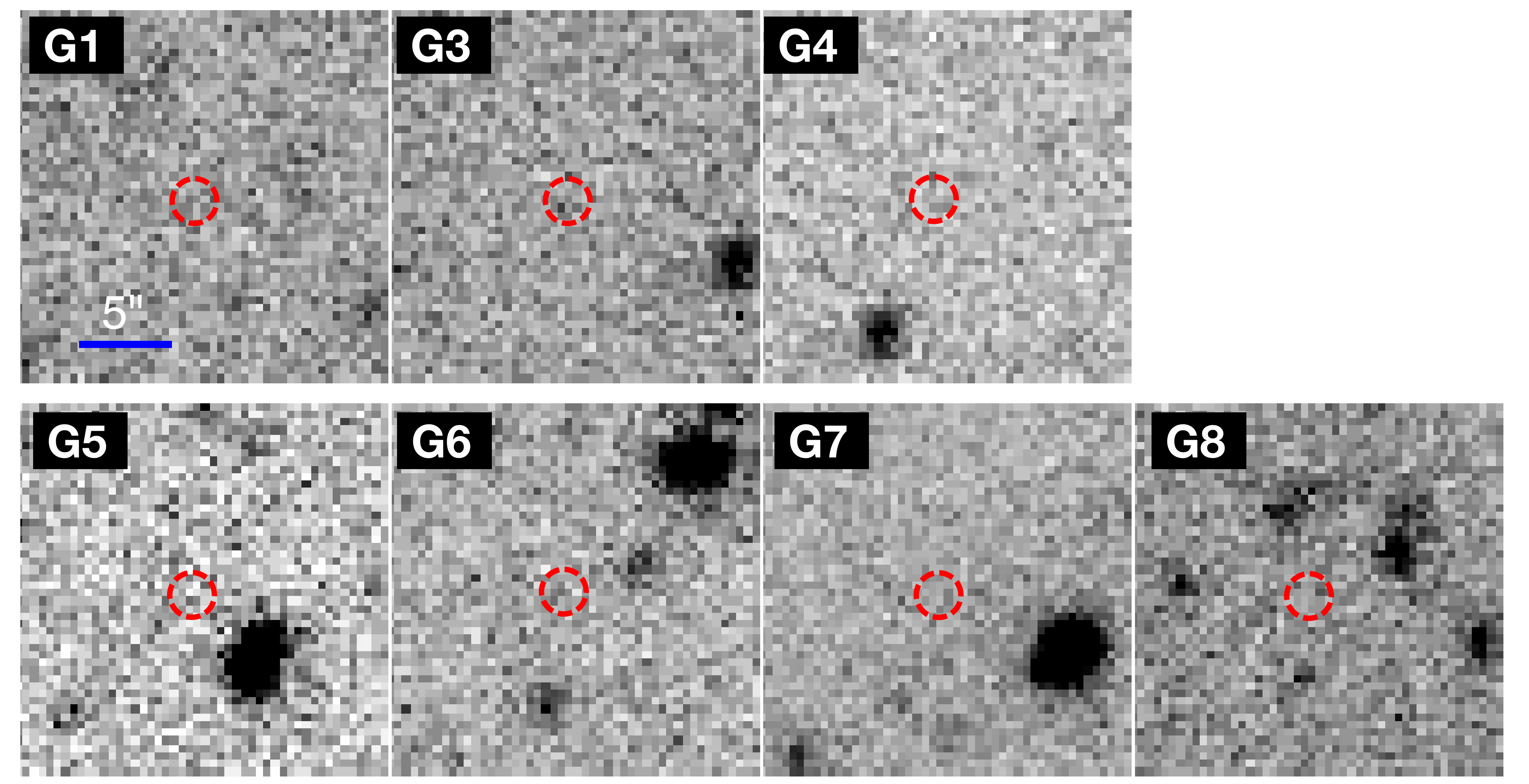}}
\caption{{\bf UVIT Postage Stamps:} Cutout (20$^{\prime\prime}$ x 20$^{\prime\prime}$) for all 7 galaxies. G1, G3, and G4 constitute the UVIT Gold sample. G1, G3, G4, G5 and G7 constitute the UVIT Silver sample. The radius of the dashed red circle in each panel is 1.2$^{\prime\prime}$.}
\label{fig:UVITstamp}
\end{flushleft}
\end{figure}

\subsection{HST and JWST observations}
\label{sec:HST}
The ultraviolet imaging used in this study is drawn from the Hubble Ultra Deep Field (HUDF) UVUDF program, a 90-orbit Treasury survey conducted with the Wide Field Camera 3 (WFC3/UVIS) aboard the \textit{Hubble Space Telescope} (Program ID GO-12534; PI: Teplitz). The UVUDF provides deep imaging in the F225W, F275W, and F336W filters; here we use only the F275W and F336W bands \citep{Teplitzetal2013, Rafelskietal2015}. In the HUDF, the observations comprise 30 orbits per filter, split into multiple exposures with typical integration times of 800–1300 s per frame, yielding total exposure times of $\sim$36 ks (F275W) and $\sim$35 ks (F336W). We use the publicly released UVUDF v2.0 mosaics, reduced with the \texttt{calwf3} pipeline and custom UVUDF software \citep{Teplitzetal2013}. The reduction includes pixel-based CTE correction, updated dark subtraction, cosmic-ray rejection, and removal of residual detector artifacts. Exposures were aligned with \texttt{TweakReg} to sub-pixel accuracy and combined using \texttt{AstroDrizzle} onto a common grid with a pixel scale of $0.06''$ per pixel. The final mosaics are well suited for accurate photometry and morphological analysis. We perform aperture photometry directly on these mosaics, applying PSF corrections based on the filter-specific curve of growth \citep{Rafelskietal2015}.

For the optical and near-infrared photometry, we use the publicly released Hubble Legacy Fields (HLF) v2.0 mosaics\footnote{\href{https://archive.stsci.edu/prepds/hlf/}{https://archive.stsci.edu/prepds/hlf/}} in GOODS–South \citep{Illingworthetal2016,Whitakeretal2019}. These include the ACS/WFC F435W, F606W, F775W, F814W, and F850LP images, as well as the WFC3/IR F105W, F125W, F140W, and F160W mosaics. The HLF products combine all ACS and WFC3/IR imaging obtained over GOODS–South from 2002–2016—incorporating data from the original GOODS \citep{Giavalisco04}, HUDF (HUDF09 \citealt{Bouwens11}; HUDF12 \citealt{Ellis13}), XDF \citep{Illingworthetal2013}, CANDELS \citep{Grogin11}, and ERS \citep{Windhorstetal2011} programs—into homogeneous, science-ready mosaics and inverse-variance weight maps. The reductions were performed using the standard \texttt{calacs} and \texttt{calwf3} pipelines, including bias subtraction, flat-fielding, dark correction, and cosmic-ray removal, followed by astrometric alignment to a common reference grid and drizzling with \texttt{AstroDrizzle} to a pixel scale of $0.06''$/pixel. Aperture photometry has been performed on the science mosaics with appropriate aperture corrections derived from empirical PSF curve of growth for each filter.

We also make use of the publicly released JWST/JADES NIRCam mosaics in the GOODS-South field, provided as High Level Science Products (HLSPs) through MAST \citep{Eisensteinetal2025,Riekeetal2023}. In particular, we use the v2.0 drizzled mosaic. The JADES HLSP reductions follow the JWST Calibration Pipeline (Stages 3) with JADES-specific enhancements for detector-level artifact removal, 1/f noise suppression, background matching, and refined astrometric solutions. The final mosaics are drizzled to a pixel scale of $0.03''$/pixel and aligned to the HST/HLF astrometric frame to an accuracy of $\sim$10–15 mas. In this study, the JADES mosaics are used solely for identifying stellar continuum counterparts to the MUSE Ly$\alpha$ detections and assessing morphological or positional offsets; no JWST photometry is used in the stacking analysis.

Where available, we further utilise archival JWST/NIRSpec low-resolution prism and medium resolution grating spectra (G140M/F070LP, G235M/F170LP: and G395M/F290LP) obtained as part of the JADES survey \citep{Bunkeretal2024,DEugenioetal2025, Curtis-Lake25} and publicly available through MAST. Each 1D JWST/NIRSpec spectrum is delivered by MAST as a Level 3 calibrated data product, following an optimal extraction procedure from the calibrated 2D spectra. The spectra are flux calibrated with a $\sim25$\% absolute flux uncertainty with the emission-lines in the grating spectra having mean flux lower by $\sim8$\% compared to the prism spectra \citep{Scholtz25}.

\begin{figure}
\rotatebox{0}{\includegraphics[width=0.5\textwidth]{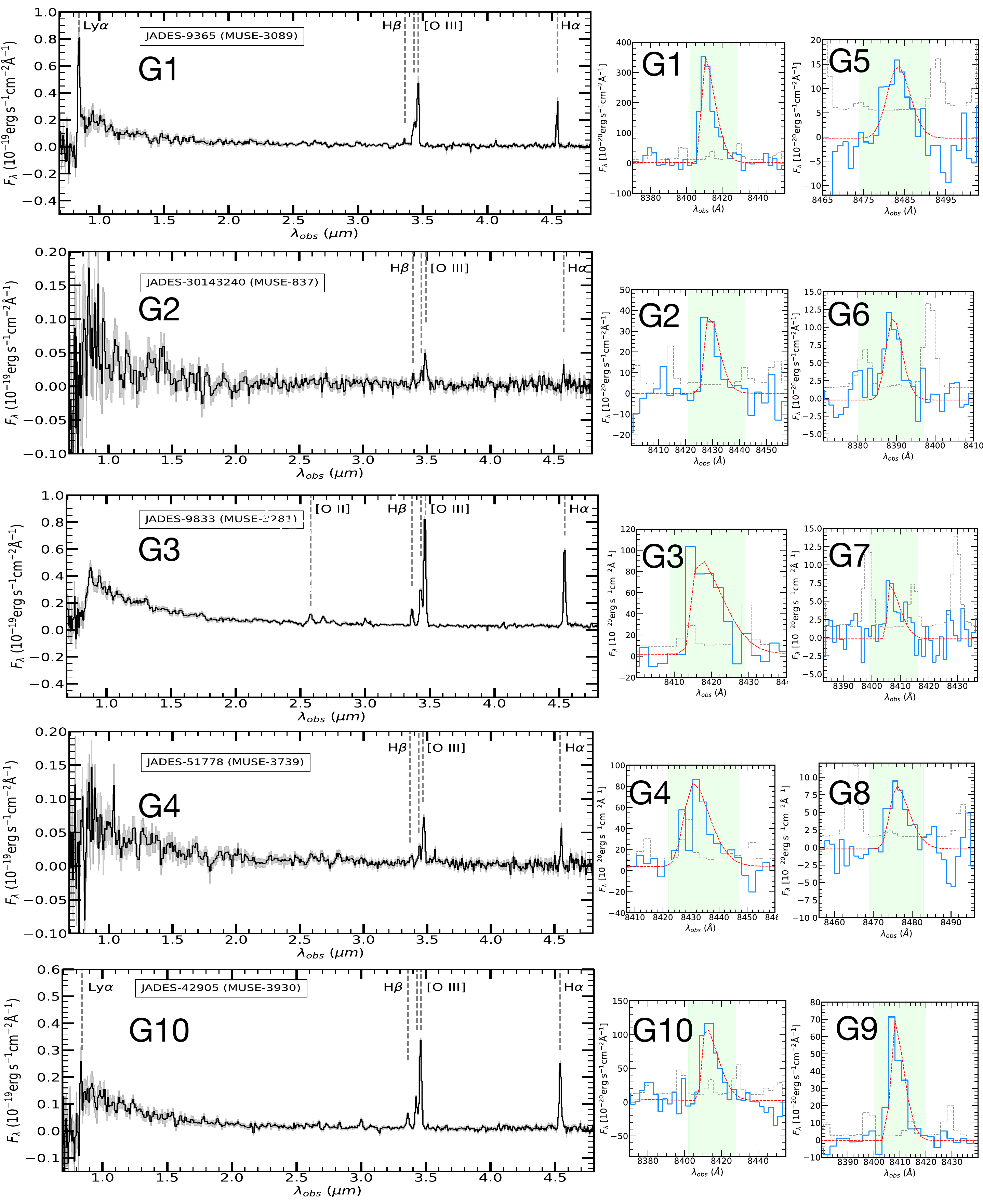}}
\caption{{\bf MUSE and JWST prism spectra}: The left panels show the JWST prism spectra of the five galaxies (G1, G2, G3, G4, G10) comprising the gold sample. For each galaxy, Ly$\rm \alpha$ emission line profile from VLT/MUSE is highlighted in the right panel.}
\label{fig:MUSEJWST}
\end{figure}

\subsection{Sample selection}
\label{sec:selection}
Our primary selection criterium is a clean and reliable detection of Ly$\alpha$ emission from galaxies within a narrow redshift range $5.9< z <6.0$ in the MUSE-HUDF. The narrow redshift range was selected to construct a homogeneous sample of Ly$\rm\alpha$ emitters (LAEs), which consequently places the Ly$\rm\alpha$ emission near the redder end of the MUSE spectral window. Notably, this redshift range also enables the study of the reionization era at a stage approaching its completion \citep{Keatingetal2020,Bosmanetal2022}. The MUSE-HUDF catalogue contains 25 such LAEs in the aforementioned redshift range.

We thoroughly re-examined each catalogued source to vet for potential low-$z$ contaminants or uncertain detections. This is done visually using a combination of the publicly available interactive catalogue browser (https://amused.univ-lyon1.fr) and the publicly released MUSE datacubes. To facilitate this, we examine subcubes of size $5^{\prime \prime}\times5^{\prime \prime}$ centered around each LAE coordinate. Additionally, we processed the MUSE datacubes with the LSDCat 2.0 software \citep{Herenz2023} to verify the MUSE-HUDF detections and to vet for potential foreground emission. In the initial vetting stage, we removed the following sources:

\begin{itemize}
    \item MUSE ID 4733 which lies near a low-redshift galaxy that is directly visible in the F606W filter, where the Lyman break would be expected.
    \item MUSE ID 7686, originally identified as a low signal-to-noise Ly$\rm\alpha$ emitter, was excluded due to its low confidence level (MUSE ZCONF = 1).
    \item MUSE ID 7970 shows a line near Ly$\rm\alpha$, close in wavelength to O~VI but redshifted by approximately 500~km/s — suggesting the Ly$\rm\alpha$ redshift might be inaccurate.
    \item MUSE ID 8264 and 8371 form a close pair of Ly$\rm\alpha$ emitters separated by 1.2 arcseconds (equal to the N242W PSF FWHM).
    \item MUSE ID 8238 due to the presence of two other sources in the UVUDF catalogue \citep{Rafelskietal2015} located $\sim 0.8^{\prime\prime}$ away.
    \item MUSE 587 was removed due to the presence of an object at a photo-z of 3.71 (possible foreground contamination) identified in the HST/UVIS/F336W filter mosaic.
\end{itemize}
This spectroscopic exercise leaves us with a sample of 19 LAEs. We further refine our sample by identifying counterparts and potential foreground contaminants using high-resolution HST and JWST mosaics.

\begin{table*}[ht]
\resizebox{\textwidth}{!}{
\begin{tabular}{cccccccccc}
\hline
Label & HST & UVIT & MUSE (NIRCam, NIRSpec) &
$L_{\mathrm{Ly}\alpha}$ &
EW(Ly$\alpha$) &
Flux$_{N242W}$ &
$\sigma_{\mathrm{pix}}$ &
SNR & $\Delta$SNR \\
& & & &
($10^{41}$ erg s$^{-1}$) &
(\AA) &
($10^{-4}$ cps) &
($10^{-5}$) & & \\
\hline
G1  & Gold   & Gold   & 3089 (142397, 9365) & $63.84\pm1.98$ & $145.7\pm44.3$ & $5.80 \pm 0.60$ & 3.88 & 2.93 & 0.30 \\
G2  & Gold   & --     & 837 (131701, 30143240) & $4.95\pm0.51$ & $89.1\pm88.1$ & -- & -- & -- & -- \\
G3  & Gold   & Gold   & 3281 (208145, 9833) & $18.10\pm1.98$ & $29.1\pm7.5$ & $5.67 \pm 1.38$ & 3.06 & 3.62 & 0.80 \\
G4  & Gold   & Gold   & 3739 (127233, 51778) & $17.9\pm1.94$ & $39.5\pm12.0$ & $5.70 \pm 0.99$ & 3.33 & 3.35 & 0.58 \\
G5  & Silver & Silver & 7974 (118814, --) & $4.04\pm1.12$ & -- & $4.73\pm0.69$ & 3.32 & 2.32 & 0.33 \\
G6  & Bronze & Bronze & 8138 (--, --) & $2.40\pm0.3$ & $77.5\pm78.6$ & $5.77 \pm 0.53$ & 3.31 & 3.41 & 0.31 \\
G7  & Silver & Silver & 8462 (117358, --) & $1.60\pm0.35$ & -- & $6.56\pm0.45$ & 4.42 & 2.91 & 0.20 \\
G8  & Bronze & Bronze & 8163 (--, --) & $2.20\pm0.33$ & -- & $1.55\pm0.78$ & 4.35 & 0.70 & 0.30 \\
G9  & Silver & --     & 6326 (295448, --) & $8.54\pm0.46$ & $176.7\pm122.3$ & -- & -- & -- & -- \\
G10 & Gold   & --     & 3930 (110739, 42905) & $22.86\pm1.77$ & $172.1\pm147.7$ & -- & -- & -- & -- \\
\hline
\end{tabular}
}
\caption{\textbf{Sample classification per instrument, catalogue cross-identification, Ly$\alpha$ luminosity and EW, Individual fluxes, SNR from UVIT.} Each object (G1–G10) is independently classified in HST and UVIT mosaics. All 10 are Bronze in HST; 7 retain the Bronze classification in UVIT. By construction, all Gold and Silver samples are subset of Bronze sample. Dashes (--) indicate cases where classification/cross identification could not be made due to lack of detection or instrument coverage. Flux from each object is measured within an aperture of radius 1.2" ($N_{pix}\sim 26$) centered on the MUSE coordinate. Col 8: $\sigma_{pix}$ refers to noise per pixel in units of $\text{cps}~\text{pix}^{-1}$). Col 9: $\text{SNR} =\text{Flux}_{N242W}/{\sqrt{N_{pix}}\sigma_{pix}}$. Error on $\text{Flux}_{N242W}$ is computed from 100 random apertures placed considering possible astrometric offset of 0.5 pix about the MUSE coordinate. $\Delta SNR$ is estimated from the random apertures fluxes due to possible offset.}
\label{tab:lae_classify_UVIT}
\end{table*}
    
Each object was visually inspected in the HST/WFC3 F160W mosaic using the DS9 software, employing a circular aperture of radius $0.5"$ centered on the coordinates extracted from the MUSE DR2 catalogue. To cross-validate potential contaminants, we examined the same regions in the JWST/NIRCam detection image mosaic from the JADES survey \citep{Eisensteinetal2025, Riekeetal2023}.
It is important to note that the JWST JADES mosaics exhibit a systematic positional offset relative to the HST UDF mosaics, primarily along Declination, with a measured shift of $\Delta{Dec}=-0.26\pm0.10$ arcsec. However, throughout this work, we adopt coordinates from the MUSE DR2 catalogue, which are aligned with the HST UDF reference frame.
The deep JWST imaging was leveraged to identify any nearby sources — particularly those undetected in the HST mosaics — located within an aperture of radius $0.5"$. This radius significantly exceeds both the astrometric offset and the instrument PSF, ensuring robust contamination assessment. 
Following this exercise, four LAEs (MUSE IDs: 304, 5215, 5875, 6523) were excluded due to potential contamination from bright neighbouring sources. An additional five LAEs (MUSE IDs: 547, 7357, 7973, 8371, and 8172) were removed following the identification of one or more nearby objects within the aperture either in the deep HST or JWST images.

This vetting process yields a final sample of 10 LAEs (see Figure~\ref{fig:HSTstamp}) suitable for image stacking in HST/UVIS imaging data. When the aperture size is increased to $1.2"$ to match the stacking analysis in UVIT/N242W, only 7 (see Figure~\ref{fig:UVITstamp}) of these 10 LAEs remain uncontaminated or free of any other source within the 1.2" aperture. 

\begin{figure*}[ht!]
\includegraphics[width = 6.in]{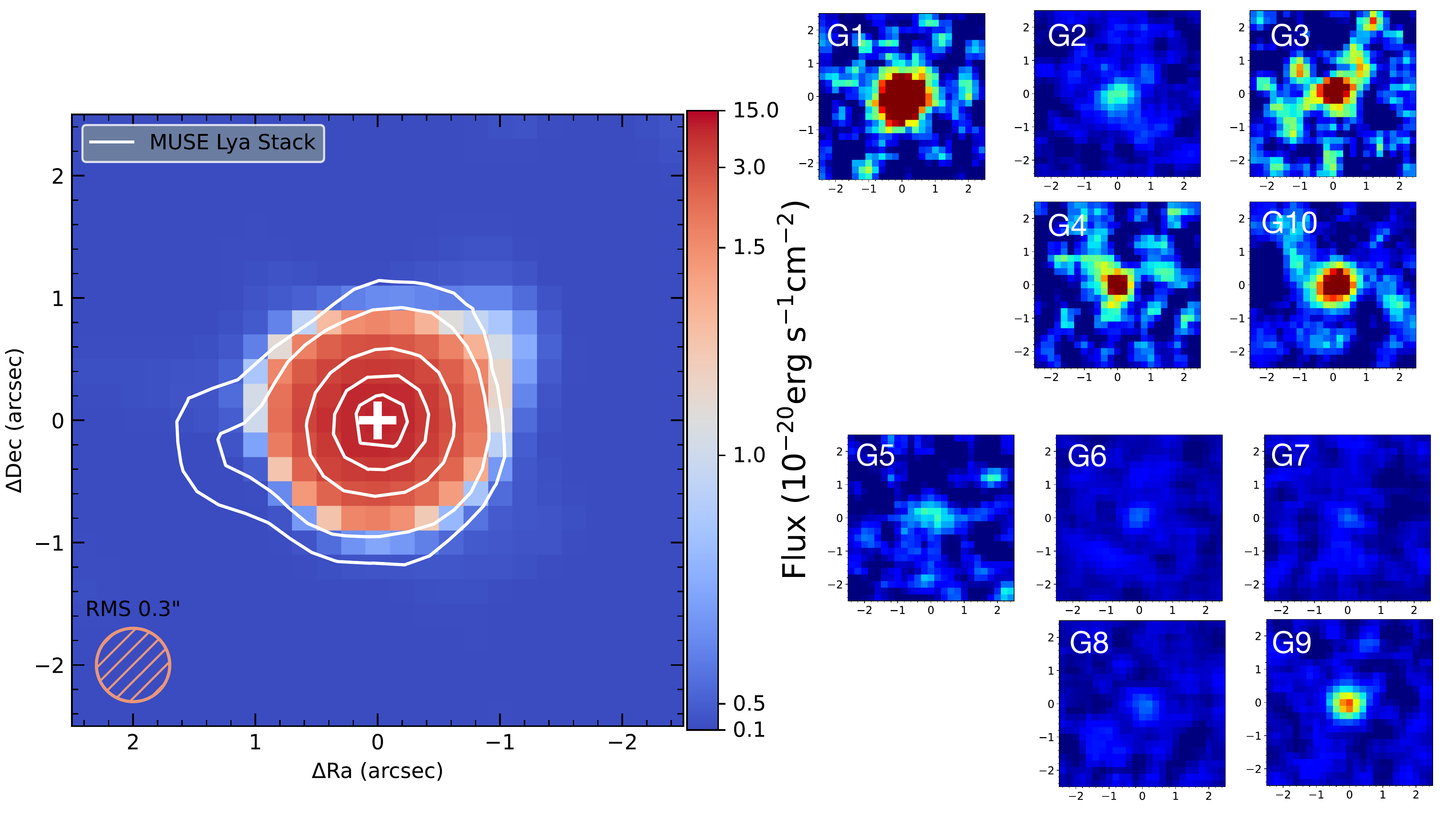}
\caption{{\bf Narrow band images of Lyman alpha emission and their stack}: Left panel shows the stacked Ly$\alpha$ narrowband image of the Bronze sample from z=5.9 - 6.0. Each LAE was aligned so that its Ly$\alpha$ centroid lies at the image center (0,0, marked by the plus sign). Right panel: Narrow band images of individual LAEs. G1, G2, G3, G4 and G10 are members of the Gold sample having JWST/NIRSpec spectra with multiple emission lines (see Figure~\ref{fig:MUSEJWST}). See Table~\ref{tab:lae_classify_UVIT} for details on individual sources.}
\label{fig:Stack_Lya}
\end{figure*}

\subsection{Sample classification}
\label{sec:classify}
It is noted that the continuum emission is not detected for all the MUSE LAEs \citep{Masedaetal2018} in the HST UV/optical or JWST/NIRCam filters. Based on the archival JWST/NIRSpec prism data \citep{Bunkeretal2024,DEugenioetal2025}, five galaxies (G1, G2, G3, G4, and G10) are found to display multiple nebular emission lines along with unambiguous non-ionizing continuum emission in the JWST filters. Based on the detection of the number of emission lines and non-ionizing continuum, we classify our sample into Bronze, Silver and Gold class:

\noindent{\textbf{Bronze criterion:}} \textit{clean Ly$\rm \alpha$ detection in the MUSE data}. All 10 LAEs (G1 - G10) constitute the HST Bronze sample, while only 7 of these (i.e., G1, G3, G4, G5, G6, G7 and G8) qualify for UVIT Bronze sample (see Table~\ref{tab:lae_classify_UVIT} for cross references).\\

\noindent{\textbf{Silver criterion:}} \textit{clean Ly$\rm \alpha$ + clear non-ionizing UV continuum detection}. There are 8 objects (G1, G2, G3, G4, G5, G7, G9 and G10) that constitute the HST Silver sample while 5 of them (G1, G3, G4, G5 and G7) belongs to UVIT Silver sample.\\

\noindent{\textbf{Gold criterion:}} \textit{ detection of Ly$\rm \alpha$ + non-ionizing UV continuum + additional emission line from JWST/NIRSpec spectroscopy}. Five out of ten (G1, G2, G3, G4 and G10) are HST Gold sample while three (G1, G3 and G4) constitute UVIT Gold sample. \\

By construction, the Gold and Silver samples are subsets of the Bronze sample. For each sample class, the corresponding sources that go in the stack of N242W imaging are also shown in the Table~\ref{tab:lae_classify_UVIT} as well as mentioned in the caption of Figure~\ref{fig:UVITstamp}. For all objects, we confirmed the absence of contaminating sources — foreground or background — within an aperture appropriate for the filters used to probe ionizing photons. This contamination assessment combined visual inspection of high-resolution JWST and HST imaging with detailed examination of the MUSE datacubes and NIRCam spectra (when available).
All galaxies in our stacking sample exhibit Ly$\rm \alpha$ emission with asymmetric line profiles characteristic of resonant scattering through an outflowing, partially neutral medium \citep[see Figure~\ref{fig:MUSEJWST},][]{Verhamme08}. 
The observed Ly$\rm \alpha$ luminosities span $L_{\rm Ly\alpha} \sim 0.16$–$6.38 \times 10^{42}$~erg~s$^{-1}$ with median of $6.8\pm0.36 \times 10^{41}$~erg~s$^{-1}$. In the four Gold sample galaxies, their Ly$\rm \alpha$ luminosities exceed the characteristic luminosity $L^* \simeq 1.0 \times 10^{42}$~erg~s$^{-1}$ at $z \sim 6$ (see Table~\ref{tab:lae_classify_UVIT}). The median (mean) EW (Ly$\alpha)=83.0\pm 40.9\ \AA$ ($104.2\pm32.6\ \AA$). Two of 10 LAEs in the sample lack detectable counterparts in deep HST and JWST imaging, consistent with low-mass, metal-poor systems exhibiting high Ly$\alpha$ equivalent widths \citep{Masedaetal2018,Masedaetal2020,Saxenaetal2024}. We are confident that the 5 other galaxies with only MUSE redshift are most certainly also z$\sim 6$ LAEs.

\begin{figure*}[ht!]
\begin{center}
\rotatebox{0}{\includegraphics[width=0.8\textwidth]{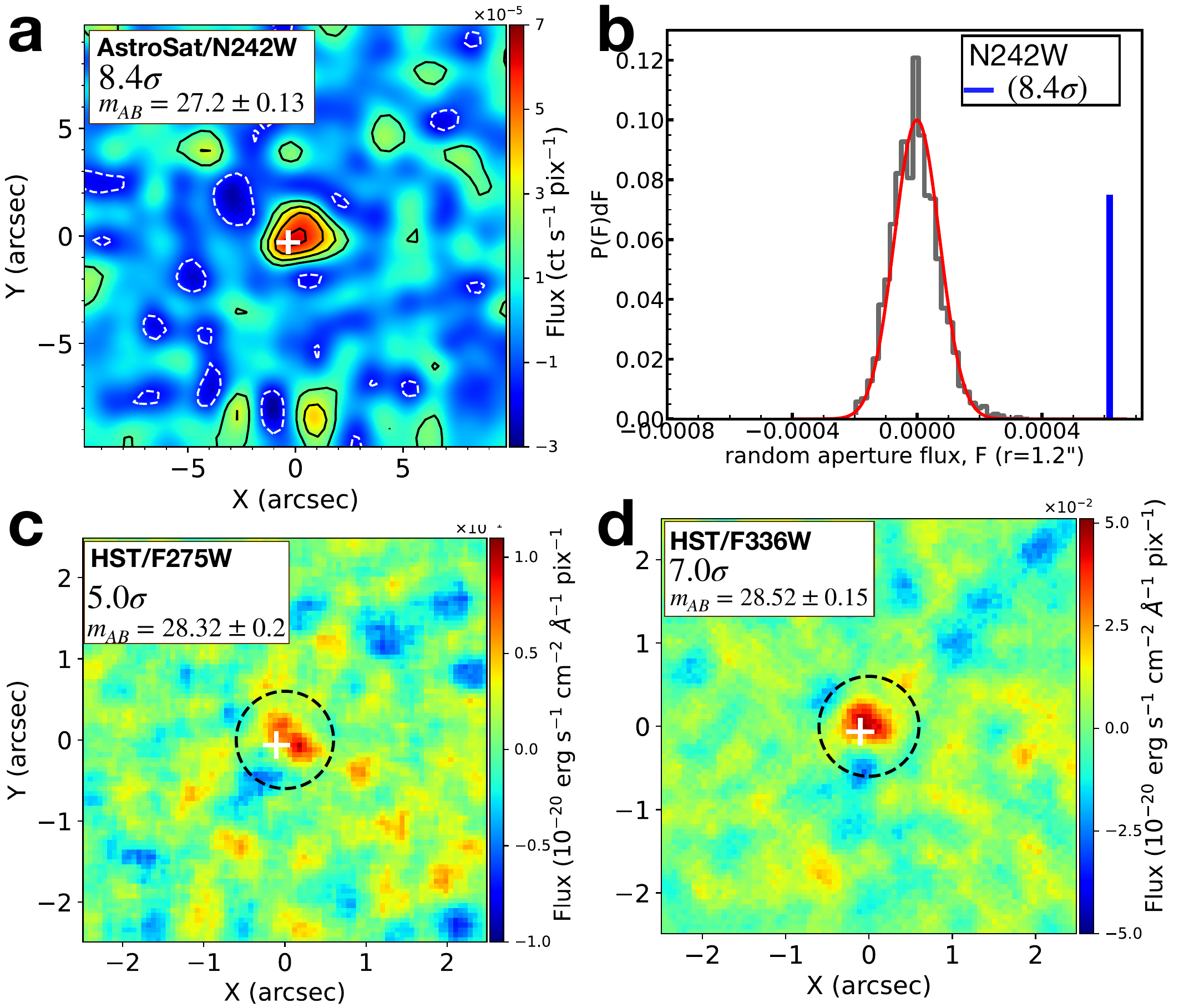}}
\end{center}
\caption{{\bf Stack Detection of ionizing photons}: Panel \textbf{a} showing the stacked image of UVIT Bronze sample emitting rest-frame $350$\AA\ emission between redshift $z = 5.9 - 6.0$ caught in the AstroSat/UVIT/N242W filter. The dark solid contours are drawn at $1\sigma$, $2\sigma$, $3\sigma$, $4\sigma$ where $\sigma = 1.44\times 10^{-5}$~ct~pix$^{-1}$~s$^{-1}$ (cps) refers to the rms of the final stacked image. The white dashed contours refer to the negative intensity contours. Panel {\textbf{b}} displays the histogram of random aperture (of radius 1.2$^{\prime\prime}$) fluxes placed on the final stacked image except the central region. The stacked flux within the same aperture from the central region is $F_{N242W} = 6.16\times 10^{-4}$~ct~s$^{-1}$($\sim 8.4 \sigma$). Panel {\textbf{c, d}} display the stacked images of HST Bronze sample in the F275W and F336W filter corresponding to rest-frame $392$\AA\ and $485$\AA\ at $z=5.94$ respectively. The radius of the dashed circle is $0.6^{\prime\prime}$. Observed magnitude and SNR of the stacked galaxies are shown in each panel; note the $\sigma$ for F275W and F336W are different from each other. The white cross marks the centroid of the non-ionizing UV continuum probed by the HST/F850LP ($\lambda_{rest}=1300\ \AA$) filter (see Figure~\ref{fig:UVcontinuum}).}
\label{fig:StackUVIT-HST}
\end{figure*}

\section{Stacking procedure}
\label{sec:stack}
We conduct the stacking analysis independently for each sample. Below, we detail the methodology employed for the N242W filter as a representative case. This procedure has consistently been applied across all filters and samples analyzed in this study.

We extract $20^{\prime\prime} \times 20^{\prime\prime}$ cutouts ($5^{\prime\prime}\times5^{\prime\prime}$ for HST filters) from the AUDF-South mosaic \citep{Sahaetal2024}, centered on the coordinates (RA, Dec, extracted from the MUSE DR2 catalogue) of the galaxies exhibiting Ly$\rm\alpha$ emission observed with VLT/MUSE \citep{Baconetal2023}. To mitigate contamination from bright neighbouring sources in the stacking analysis, we mask all detected objects outside a circular aperture of radius $1.2^{\prime\prime}$ (or $0.3^{\prime\prime}$ for HST cutouts), preserving only the pixels within the central aperture. Our choice of a more conservative aperture size of $0.3^{\prime\prime}$ for HST filters, compared to $0.5^{\prime\prime}$ adopted in previous studies \citep{Smithetal2018, Wangetal2025}, reduces the effective sky area by a factor of $\sim (0.5/0.3)^2 \simeq 3$ and thereby decreasing the likelihood of contamination from unrelated line-of-sight sources. Object detection and masking are performed using SExtractor \citep{Bertin96}, with a detection threshold of $0.8\sigma$ and DETECT\_MINAREA=6 for N242W, optimized to identify sources while accounting for the extremely low surface brightness \citep{Merlinetal2024} of the LyC-emitting regions (typically fainter than 27 AB mag).
This masking strategy effectively suppresses potential contamination from foreground, non-ionizing sources close to the line of sight for the target of interest \citep{Sianaetal2010,Rutkowskietal2016}. We note that on average $\sim 20\%$ of the total pixels are masked in each cutout image of the N242W filter. Around each masked region, we define an annuli to denote neighbouring region. Each of the masked regions are then filled by randomly drawing unsegmented pixels from the neighbouring region, analogous to the fixpix routine in IRAF. Due to the randomness involved in filling the masked regions, each iteration produces a different background RMS while the central aperture pixels remain unmodulated. The essence of the stacking procedure presented here is similar to that of \cite{MaulickSaha_stack2026}, who applied it to stack a set of galaxies at $z\sim 1$ using the F154W filter of UVIT. We acknowledge that this approach may exclude faint, extended, or off-centered LyC emission (beyond $0.3^{\prime\prime}$ for HST and $1.2^{\prime\prime}$ for AstroSat/N242W, corresponding to physical scales of 2 – 7 kpc). However, given the small number of sources in each stack, we manually inspect each cutout to confirm the absence of nearby contaminating objects. The choice of a conservative aperture radius minimizes the likelihood of including random, low-redshift interlopers \citep[see Section~\ref{sec:contamination} below, ][]{Smithetal2018}.

Prior to stacking, we estimate and subtract the residual sky background from each individual cutout. This is achieved by placing random apertures — matched in size to the full width at half maximum (FWHM) of the PSF for the corresponding filter (here N242W) — across the cutout, and computing the distribution of enclosed fluxes. The resulting histogram is fitted with a Gaussian profile; the mean (or mode, in case of skewed histogram) is interpreted as the residual background level per pixel and subtracted from the entire cutout. This procedure effectively centers the sky background distribution around zero, mitigating local background variations across the mosaic \citep{Smithetal2018}. Such variations can arise from instrumental effects including imperfect flat-fielding or background subtraction, particularly in low-photon-count regimes such as in the UV \citep{Pandey25}. We perform both mean and median stacking of the residual background-subtracted, masked cutouts. The central aperture flux is measured within a circular aperture of radius $1.2^{\prime\prime}$ for the N242W filter centered on the stacked image. To estimate the significance of the detected signal, 200 randomly placed apertures of the same size are positioned across the background region of the stacked image (avoiding the central aperture) to characterize the noise distribution. {Given the stochastic nature of pixel replenishment process in the masked regions (explained above), the entire stacking procedure is repeated 50 times. Each iteration generates a unique realization of the stacked image, yielding a total of 10000 randomly placed background apertures across all runs in total}. The fluxes from these background apertures are then used to construct a histogram that is well-described by a Gaussian profile (see Figure~\ref{fig:StackUVIT-HST}b, for example). We fit this distribution to determine the mean (or mode, when skewed) and standard deviation which quantifies the noise level in the stacked image. For each filter, the stack flux is aperture-corrected based on its own PSF encircled energy curve.   

\begin{figure*}
\begin{center}
\rotatebox{0}{\includegraphics[width=0.8\textwidth]{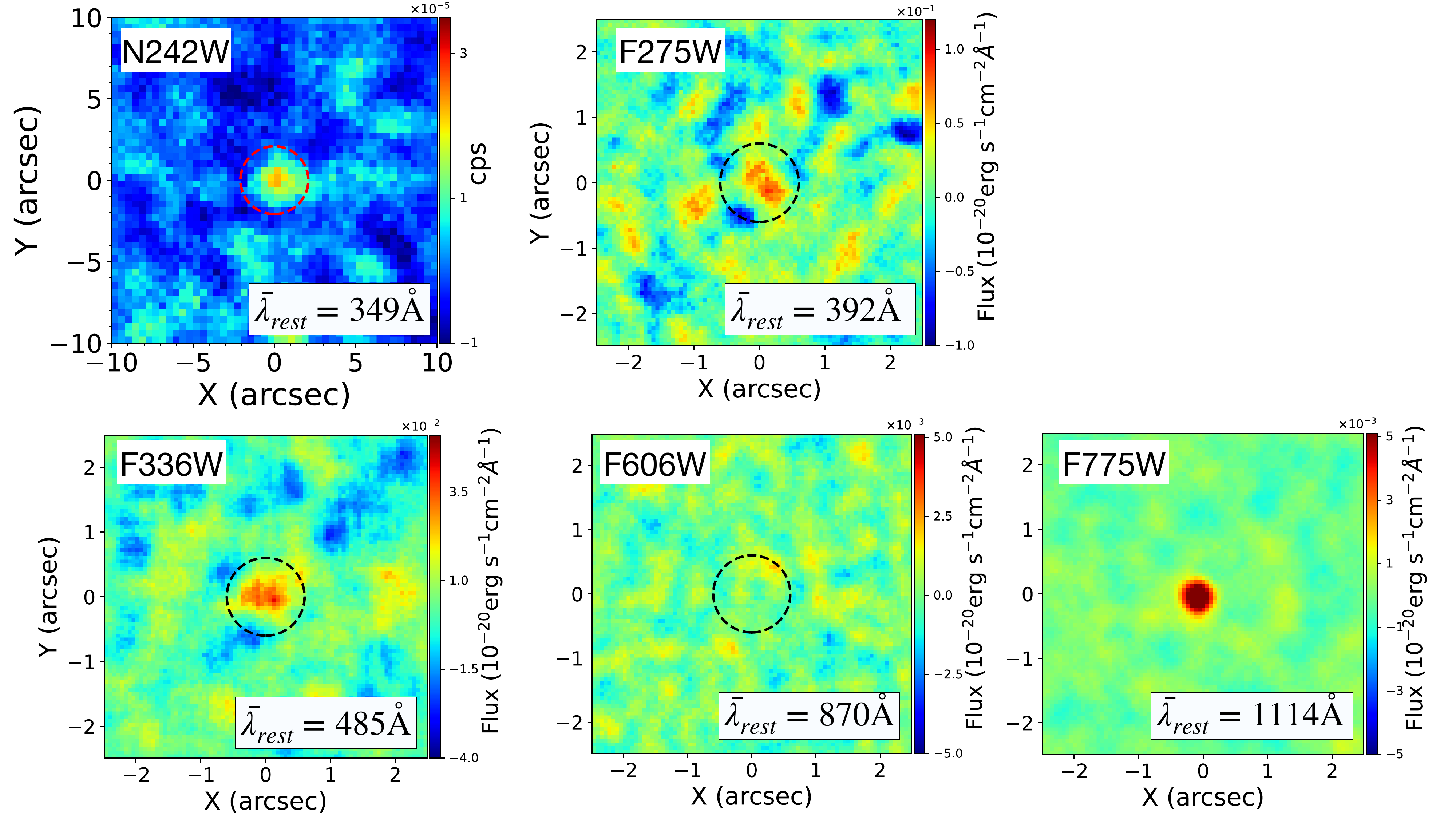}}
\caption{{\bf Gold stack:} Panels showing stacked images considering only the Gold sample galaxies (G1, G2, G3, G4 and G10). For each stacked image, the color scheme is same but color-bar scale has been customized. Each image has been smoothed with a tophat kernel. The rest-frame wavelength corresponds to the filter mean wavelength after redshifting.}
\label{fig:Goldstack}
\end{center}
\end{figure*}

An identical stacking procedure — including source detection, masking, and background estimation — was applied to HST UV bands WFC3/UVIS F275W and F336W from the UVUDF survey \citep{Rafelskietal2015}. In addition, we stack deep HST optical and near-infrared imaging data \citep{Illingworthetal2013} from ACS/F435W, F606W, F775W, F814W, F850LP and WFC3/IR F105W and F125W. These additional stacks are used to probe ionzing spectrum above $\sim 500 \AA$ and the UV continuum from non-ionizing photons (see Table~\ref{tab:photometry_stack}). For the HST F275W and F336W stacks, the fraction of masked pixels ranges from 8–15\%, indicating that masking was conservative and did not artificially alter the background RMS. For the stack fluxes and magnitudes in all the filters explored in this study, the readers are referred to Table~\ref{tab:photometry_stack}. 

We also applied this identical stacking methodology to MUSE Ly$\alpha$ narrow-band images and to broad-band slices extracted from the MUSE datacube spanning $\lambda_{\rm obs} = 4859$–$5522\AA$ (corresponding to rest-frame $700$–$800 \AA$ at z=5.94) for the HST Bronze sample. Figure~\ref{fig:Stack_Lya} shows the individual narrow-band images of the LAEs (G1 - G10) and their stacked signal with a strong detection at $22\sigma$ (where $\sigma = 8.6\times10^{-20}$~erg s$^{-1}$cm$^{-2}$). The Ly$\alpha$ stacked image has been convolved with a Gaussian kernel with $\sigma_{diff}^2 = \sigma_{UVIT}^2 - \sigma_{NB}^2$ to compare with the N242W filter stack.

\section{Stack detection of ionizing photons along the Lyman Valley}
\label{sec:detection}

In this section, we present our stacking detection of ionizing photons for $\lambda < 912 \AA$. At the mean redshift of z=5.94, this part of the ionizing spectrum is probed by HST filters F606W (partially), F435W, F336W, F275W and AstroSat/N242W. First, we focus on the wavelength region $\lambda < 504$\AA, marking the HeI ionization edge and then explore the other wavelength region with $\lambda > 504$\AA. We choose this division based on our analysis of the Lyman Valley transmission which suggests a non-zero transmission below the He{\sc i} ionization edge at the mean redshift ($z\sim 5.94$) of the stack (see section~\ref{sec:IGM}). 

\subsection{Detection of ionizing photons ($\lambda_{rest} < 504 \AA$)}

Figure~\ref{fig:StackUVIT-HST}(a) shows the direct detection of ionizing photons at rest-frame wavelength of $350\text{\AA}$ from a stack of LAEs in the UVIT/N242W Bronze sample, with a signal detected at $8.4\sigma$ above the background. We place random apertures (of size $1.2^{\prime\prime}$) on the final stacked image (except the central region) to construct a probability distribution of these random aperture fluxes by fitting a Gaussian function (see section~\ref{sec:stack} for details), see Figure~\ref{fig:StackUVIT-HST}(b) for the N242W stack. We also utilize this to estimate the noise from the final stacked image. From the fitting, we obtain a $\sigma_{aper} = 7.26\times10^{-5}$ which translates to a $\sigma = \sigma/\sqrt{N_{pix}} \simeq 1.42\times10^{-5}$~ct~pix$^{-1}$~s$^{-1}$ ($N_{pix}=26$). One can easily see that $\sigma \simeq \sigma_{pix}/\sqrt{7} \simeq 1.4\times10^{-5}$ (see Table~\ref{tab:lae_classify_UVIT}). In other words, the noise in the final stacked image scales as $1/\sqrt{N}$, where N is the number of galaxies in the stack, indicating that it approximately follows uncorrelated Gaussian statistics. None of the 10000 random apertures yield a signal comparable to the measured SNR$\sim 8.4\sigma$, implying a false detection probability $P < 10^{-4}$.
Stacking the HST Bronze sample yields corresponding detections at rest-frame $392\text{\AA}$ and $485\text{\AA}$ in the WFC3/F275W and F336W filters, with significances of $5.0\sigma$ and $7.0\sigma$, respectively (Figure \ref{fig:StackUVIT-HST} c,d) and a false detection probability $P < 10^{-4}$ based on 10000 random apertures. The detection of ionizing photons in three filters thereby probe the rising part of the Lyman Valley transmission ($\lambda < 504 \AA$). The Galactic extinction and aperture corrected magnitude in the N242W, F275W and F336W filters are $27.16\pm 0.13$, $28.32\pm 0.2$ and $28.52\pm0.15$ AB mag respectively, with corresponding fluxes listed in Table~\ref{tab:photometry_stack}. 

Detection significances for the Gold sample are $6.1\sigma$, $4.9\sigma$, and $4.5\sigma$, in N242W, F275W and F336W respectively (see Figure~\ref{fig:Goldstack}). The same for the Silver sample are presented in Table~\ref{tab:photometry_stack}. Although these subsets are smaller in terms of sample size, and yield a relatively lower SNR, the detection remains statistically significant even under progressively stricter selection criteria. It is worth noting that the morphology of the stack in the Bronze and Gold samples are qualitatively similar in their respective filters. This indicates that the Gold selection does not introduce a measurable bias in the spatial distribution of the ionizing radiation in their respective filters. The agreement between the full sample (Bronze) and the higher-quality subset (Gold) further demonstrates that the stacked signal is robust against subsampling and is not driven by a small or atypical subset of galaxies.
Before we discuss further the morphology of the stacked images, we report non-detection in the F606W (having the deepest HST imaging in the GOODS-South Field) filter stack (probing $\lambda_{rest} = 870\AA$) and a clean detection in F775W filter representing the non-ionizing UV continuum at $\lambda_{rest} = 1114\AA$ (see Figure~\ref{fig:Goldstack}).

\begin{figure*}
\begin{center}
\rotatebox{0}{\includegraphics[width=0.8\textwidth]{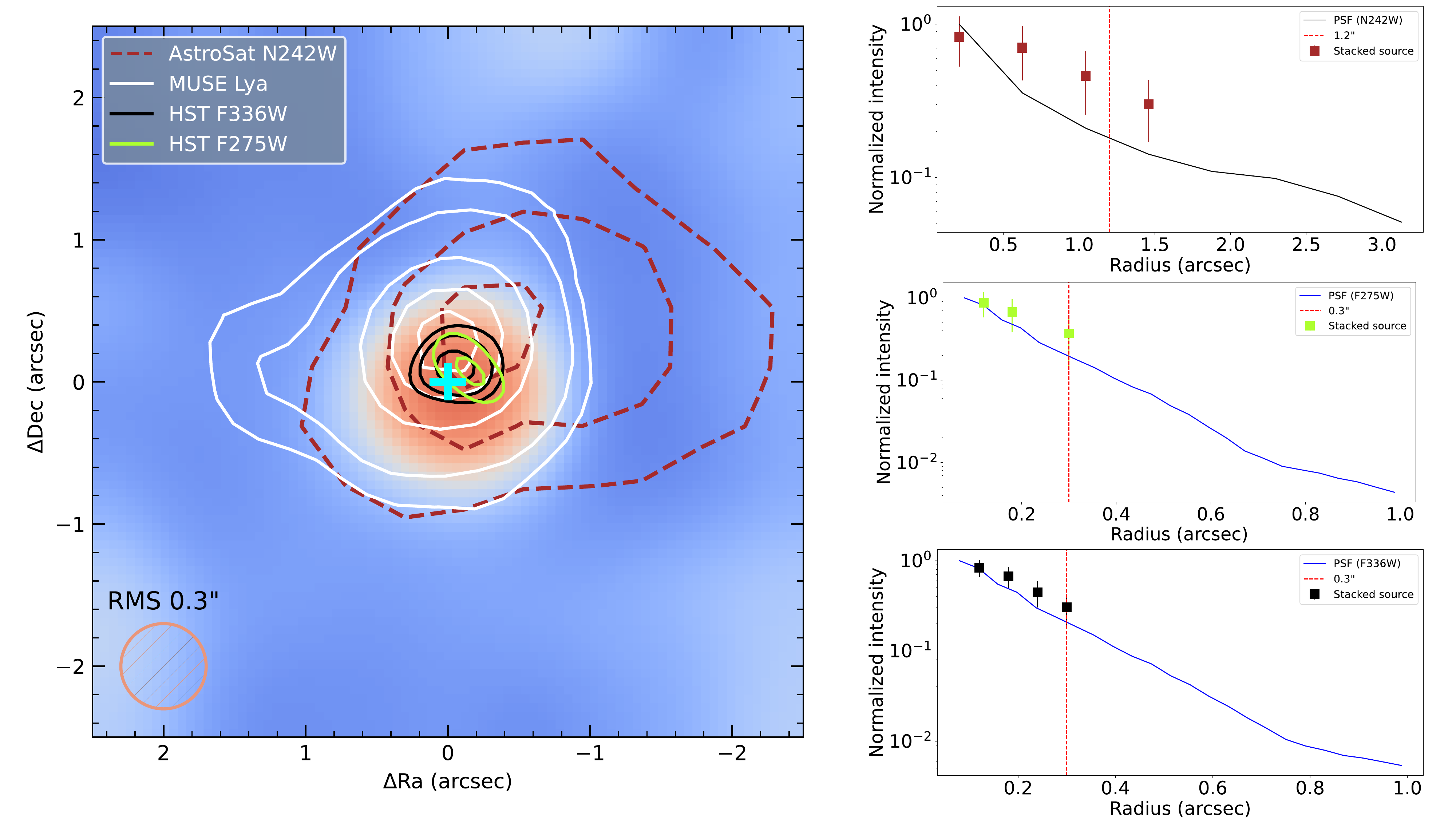}}
\caption{Left panel: Contours of Ly$\rm\alpha$ and ionizing photon fluxes from three filters overlayed on the stack UV continuum measured in HST/F850LP. Right panel: Stack surface brightness profiles (normalized) in each filter are compared with their respective PSF profile. Red dashed line in each panel marks the aperture radius used for calculating the stack flux.}
\label{fig:contoursPSF}
\end{center}
\end{figure*}

\subsection{Offset and Morphology}
\label{sec:stackmorph}

It has been observed for some galaxies at redshift $z\gtrsim1$, that the Lyman continuum emission is offset with respect to the non-ionizing UV continuum \citep{Ji20,Dhiwar24,Gupta24,Yuan24,Maulick25,RiveraThorsen26}. To investigate this, we estimate the light-weighted centroid of the non–ionizing UV continuum using the F850LP stack, which probes the rest-frame $\sim1300\AA$ (see left panel of Figure~\ref{fig:contoursPSF}). The centroid is computed within a small box around the approximate center using a center-of-light (flux-weighted) method appropriate for faint stacked signal. We apply the same procedure to the F336W, F275W, and N242W stacks to determine their relative centroid positions. The angular separation between the F850LP and F336W centroids is $0.036"$, while the offsets between F850LP and F275W and between F850LP and N242W are $0.075"$ and $0.5"$ (typically, range from sub-kpc to few kpc at $z\simeq 6$), respectively. Offsets between LyC emission and the UV continuum on kpc scales have been reported at $z\sim 3$ \citep{Gupta24}, although such detections remain observationally challenging. In our stacks, the contours of the F336W and F275W images are nearly coincident with the non-ionizing UV continuum, showing no significant spatial displacement. In contrast, the Ly$\alpha$ and N242W stacks (the latter having a larger PSF) exhibit apparent offsets with respect to the HST/F850LP centroid. Spatial offsets between Ly$\alpha$ emission and the UV continuum are commonly observed at high redshift ($z\sim 3 – 7$), typically on kpc scales and affecting a substantial fraction of galaxies \citep{Lemauxetal2021}. While there is a mild trend of increasing displacement toward shorter wavelengths, these offsets are not statistically significant, as their amplitudes are smaller than the PSF FWHM and comparable to the uncertainties in astrometry and centroid determination.

To assess the intrinsic morphology of our stacked detections at these short wavelengths, we compared the stack surface-brightness (SB) profiles with the corresponding PSF profiles in each filter. PSFs for the WFC3/UVIS F336W and F275W filters were downloaded from the STScI website\footnote{\url{https://www.stsci.edu/hst/instrumentation/wfc3/data-analysis/psf}}. An image cube of 56 PSFs per filter was visually examined. The SB profiles of the PSFs and stacks were extracted using concentric circular apertures centered on their respective centroids. The stack SB profiles were measured with 1$\sigma$ uncertainties to assess consistency with the PSF. As shown in Figure~\ref{fig:contoursPSF}, the stacked SB profiles in both F336W and F275W closely follow the PSF profiles out to $\sim 0.3"$ radius. Similarly, the N242W stack follows its PSF out to $\sim1.2"$ (see right panel of Figure~\ref{fig:contoursPSF}). Beyond these radii, the stack primarily samples noise, as evident in the stacked images in all the filters. Within the 1$\sigma$ uncertainties, the stacks are consistent with their respective PSF, indicating that the detected emission is unlikely to be spatially extended. Visual inspection of the PSF light distribution reveals that the UVIS PSFs (especially, the core part) are slightly elongated along the upper-left to lower-right diagonal (e.g., PSF number 7 in the cube), consistent with the WFC3/UVIS astigmatism \citep{Sabbi2009,SabbiBellini2013}. This elongation arises from the tilt of the WFC3/UVIS channel and becomes prominent when the telescope is slightly off optimal focus. It is therefore likely that the apparent morphology of our UVIS stacks reflects the intrinsic PSF shape rather than any genuine extension of the sources.

\subsection{Detection of ionizing photons ($\lambda_{rest} > 504 \AA$)}

\begin{figure*}
\begin{center}
\rotatebox{0}{\includegraphics[width=1\textwidth]{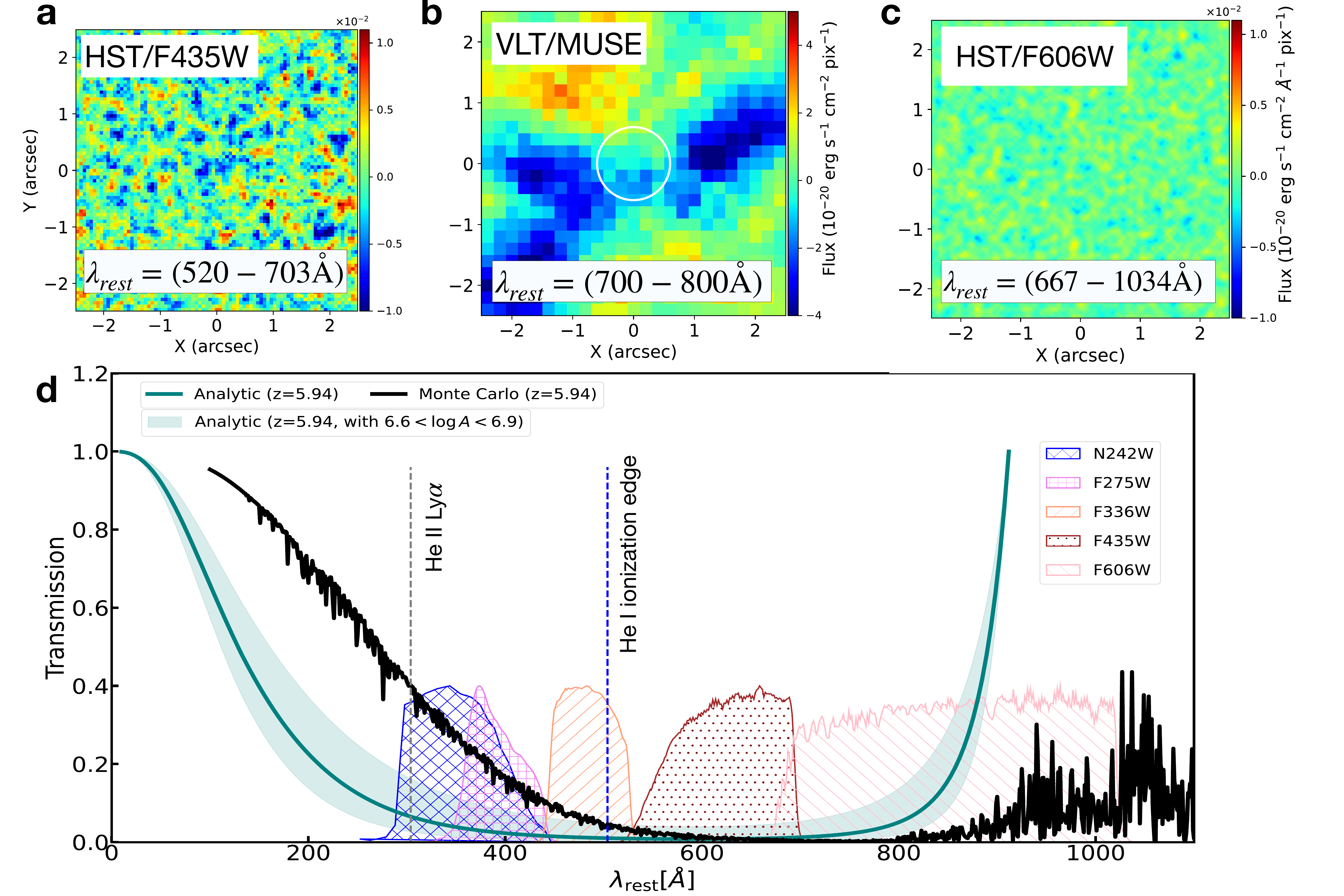}}
\caption{\textbf{Lyman valley, Non-detection of ionizing continuum from $\lambda_{rest}=500 - 800\AA$:} Panel \textbf{a}: Stack of Bronze sample in the ACS/WFC/F435W filter (covering the rest-frame wavelength of $520 - 703\AA$ of HST showing null detection.  Panel \textbf{b}: Stack of MUSE sub-CUBE with spatial dimension $5"\times5"$ (same as in panel a) - the flux within the solid circle of radius 3 pix is $-0.24\times10^{-18}$~erg s$^{-1}$~cm$^{-2}$. The $3\sigma$ upper limit is given by $1.9\times10^{-18}$~erg s$^{-1}$~cm$^{-2}$.Panel \textbf{c}: Same as in Panel \textbf{a} but for the F606W mosaic image with the deepest HST observations. Panel \textbf{d}: The Lyman Valley (LV) transmission at $z=5.94$, the mean redshift of the stack, from a simple analytic model of photoelectric optical depth with a neutral hydrogen column density function in the range of $11 \le \log N_{HI}\le20$ (see Section~\ref{sec:IGM}) is shown by the solid curve (teal colour). The shaded area represents a 0.5-dex variation in the number density of the hydrogen absorbers. The most optimistic transmission at $z=5.94$ from extensive Monte Carlo simulations of the intergalactic transmission is shown by the black curve (see Section~\ref{sec:IGM}). These models show that the LV transmission is nearly zero in the range of $\lambda_{rest} = \sim 500 - 850\AA$, while non-zero transmission is possible at $\lambda_{rest} < 504\AA$.}
\label{fig:LVtransmit}
\end{center}
\end{figure*}

To further probe the Lyman valley in the rest-frame wavelength range $\sim 500 - 912\AA$ (Figure~\ref{fig:LVtransmit}), we stacked HST/ACS F435W filter images and the medium-band images extracted from MUSE cube (from $\lambda_{obs} = 4859 - 5522 \AA$) for the HST Bronze sample. The F435W filter probes rest-frame wavelengths of $520-703\AA$, while the MUSE samples $700 - 800 \AA$ — together covering the bulk of the expected trough in Lyman valley transmission at $z=5.94$ (see Figure~\ref{fig:LVtransmit}d and section~\ref{sec:IGM}). We additionally stacked {the HST/ACS F606W mosaic - the deepest HST data available} - which covers rest-frame wavelengths both shortward and longward of the redshifted Lyman limit ($6329 \AA$ for $z=5.94$). No significant flux is detected in the central apertures in any of these bands (Figure~\ref{fig:LVtransmit}a,b,c). However, redward of the redshifted Lyman limit, clean detections of non-ionizing UV continuum photons in HST/ACS/F775W (probing $\lambda_{rest} = 980 - 1243\ \AA$) and F814W (probing $\lambda_{rest} = 990 - 1388\ \AA$) are shown in Figure~\ref{fig:LVtransmit}(e,f) with stack fluxes and SNR presented in Table~\ref{tab:photometry_stack}. The non-detections in the rest-frame $520 - 800$ \AA\ range, jointly with the robust detections at shorter rest-frame wavelengths ($350 - 485\ \AA$) reinforce the simple theoretical expectation of high-z IGM transmission (Figure~\ref{fig:LVtransmit}). The absence of any significant flux in the stack of F606W mosaic - the deepest available HST optical observation \citep{Illingworthetal2013} - provides strong evidence against any substantial contamination from low-redshift interlopers. This has been verified in both the Bronze and Gold sample stacks independently (Figure~\ref{fig:Goldstack}). Further, the F606W band non-detection indicates the sharp decline of IGM transmission at $z\sim 6$, where the mean free path of $900 \AA$ photons falls below $\sim 1 - 2$ pMpc due to the high incidence of Lyman limit systems which have an optical depth $>1$ around the Lyman limit and dominate the IGM opacity \citep{Worsecketal2014,Beckeretal2021}. This behaviour is well reproduced in our Monte Carlo (MC) simulations of IGM transmission (see section~\ref{sec:IGM}). On the other hand, in the rest-frame wavelength below $504 \AA$, the IGM transmission actually reascends due to the smaller photoelectric cross section at shorter wavelengths. In those wavelengths, the higher column density absorbers like sub-DLAs and DLAs exceed the optical depth of unity and dominate the IGM opacity. Our MC simulations show that the most optimistic case assuming less abundant sub-DLAs/DLAs than the mean IGM allows a modest but non-negligible transmission in the rest-frame wavelength below $504 \AA$. The F336W detection, probing rest-frame $485 \AA$, lies at the edge of this recovery of transmission and is only reproduced along the most favourable sightlines, though rare (see section~\ref{sec:IGM} for details). 


\begin{table*}[ht]
\centering
\caption{Photometric measurements and detection significance of the stack in the ionizing and non-ionizing filters.}
\begin{tabular}{|l|c|c|c|c|c|}
\hline
\textbf{Filter} & \textbf{mag (Bronze)} & \textbf{Flux Density (Bronze)} & \textbf{SNR (Bronze)} & \textbf{SNR (Silver)} & \textbf{SNR (Gold)} \\
\hline
 & (AB)& (10$^{-20}$ erg s$^{-1}$cm$^{-2}${\AA}$^{-1}$) & & &\\
\hline
N242W   & $27.16 \pm 0.13$     & \texttt{25.60} $\pm$ \texttt{1.30}     & 8.4  & 6.22  & 6.14 \\
F275W   & $28.32 \pm 0.2$     & \texttt{6.92} $\pm$ \texttt{0.55}     & 5.0  & 4.62  & 4.9 \\
F336W   & $28.52 \pm 0.15$     & \texttt{3.74} $\pm$ \texttt{0.23}     & 7.0  & 5.15  & 4.5 \\
F435W   & $>31.0$ (3$\sigma$)  & $<$\texttt{$1.7 \times 10^{-21}$}                      & $<$1.0 & --   & --  \\
F606W   & $>31.5$ (3$\sigma$) & $<$\texttt{$7.3 \times 10^{-22}$}                      & $<$1.0 & --   & --  \\
F775W   & $29.74 \pm 0.12$     & \texttt{0.232} $\pm$ \texttt{0.01}     & 10.45  & --   & --  \\
F814W   & $28.75 \pm 0.14$     & \texttt{0.523} $\pm$ \texttt{0.029}     & 7.83  & --   & --  \\
F850LP  & $27.72 \pm 0.038$    & \texttt{1.02} $\pm$ \texttt{0.014}     & 30.8 & --   & --  \\
F105W  & $27.91 \pm 0.044$    & \texttt{0.664} $\pm$ \texttt{0.012}     & 24.9 & --   & --  \\
F125W  & $27.93 \pm 0.05$    & \texttt{0.50} $\pm$ \texttt{0.010}     & 20.38 & --   & --  \\  
F140W  & $27.89 \pm 0.12$    & \texttt{0.42} $\pm$ \texttt{0.024}     & 8.91 & --   & --  \\
\hline
\end{tabular}
\label{tab:photometry_stack}
\end{table*}

\section{Robustness of the stacked signal}
\label{sec:robust}

Our detections of ionizing photon fluxes are statistically significant and unlikely to result from some random background fluctuations or spurious sources/noise peaks or even boosted by a particular source (e.g., an outlier) in the sample. In the following, we perform a number of statistical tests to validate our stacking results.

\subsection{Individual fluxes, SNR and stacking of UVIT sources}

In Table~\ref{tab:lae_classify_UVIT}, we present the fluxes and their SNR for individual sources in the UVIT sample within an aperture of size 1.2" ($N_{pix} \simeq 26$). We estimate error on the fluxes by considering possible astrometric offset (with an RMS $\Delta r \sim 0.5 - 0.6$ pixel) and by placing N random apertures around the MUSE coordinate. Now, it is possible to utilize these fluxes and noise within the apertures, to estimate significance of our stacking in the following way:

\begin{equation}
    F_{stack} = \frac{\Sigma{w_j F_j}}{\Sigma{w_j}}; \ \sigma_{stack}=\frac{1}{\sqrt{\Sigma{w_j}}}, 
\end{equation}

\noindent where $w_j = 1/\sigma_j^2$ refers to the inverse variance within the aperture. One can write $\sigma_{j} = \sqrt{N_{pix}} \sigma_{pix}$, $\sigma_{pix}$ being the noise per pixel (note: since UVIT pixels are not drizzled, correlated noise per pixel is negligible, \cite{Sahaetal2024}) Nevertheless, we always compute $\sigma_{j}$ for a given aperture directly from the image. Considering the UVIT Gold sample, we obtain an SNR $\sim 5.7$ which is about 7\% less than the quoted SNR directly measured from stacking the images. If we use the rest of the four sources (G5, G6, G7 and G8), we obtain an SNR $\sim 4.8$. Whereas if we use all the sources, we obtain an SNR $\sim 7.5$. Now, if we consider the impact of possible astrometric offset and model it as a Gaussian profile:

\begin{equation}
    F_{measured} = F_{true} e^{- \frac{{\Delta r}^2}{2 \sigma_{PSF}^2}}, 
\end{equation}

\noindent where for the UVIT/N242W filter, $\sigma_{PSF} = FWHM/2\sqrt{2ln2} = 1.22$ pixel. If we plug in the numbers, we obtain $F_{measured} \simeq 0.9 F_{true}$. So in a statistical sense, the measured fluxes are underestimated by about $10\%$ from the true fluxes due to possible astrometric offset. If this is true, we expect an SNR $\sim 8.25$ when all objects are stacked. Adopting a similar argument, we would expect the SNR of the Gold sample only to be $6.3$ (against what we measure directly from the image stack as 6.1). In conclusion, starting from these individual fluxes and SNR measurements, it is possible to estimate analytically the SNR of any stacked sub-sample of UVIT. We have checked that these estimates agree with the direct SNR measurements on the stacked images presented in  section~\ref{sec:detection} and Table~\ref{tab:photometry_stack}.    

\subsection{Blank-Sky Null Test}

To quantify the likelihood that our detected LyC signal is a spurious result of faint undetected sources or false detection probabilities, we perform a null test using a set of cutouts from blank sky locations in each HST mosaic. These locations are carefully selected to be devoid of known sources (at the centre of the cutouts) based on deep HST UVUDF mosaic \citep{Rafelskietal2015}, AUDF-south in N242W \citep{Sahaetal2024}, and JWST detection images \citep{Eisensteinetal2025, Riekeetal2023}, and verified to be free of any sources in MUSE data \citep{Baconetal2023}. We choose 60 such blank-sky positions and create cutouts appropriate for each filter (same size as it was for galaxy stacking). From these 60 cutouts, we randomly draw a subset of cutouts at a time — matching the number of sources in our LAE sample — and apply an identical stacking procedure including background subtraction and masking (with identical parameters used in the case of Bronze sample stacking for each filter) to each subset. It is worth reminding here that the Bronze stack has 7 galaxies in N242W filter and 10 galaxies in the HST filters. During each stack, we compute the flux from the central aperture which remained unaltered during the stacking procedure as it was the case in galaxy stacking. We place 200 apertures (same size as the central one used for flux calculation e.g., 0.3" for HST filter) randomly on the stacked image and measure fluxes within. The histogram of these fluxes are fit with a Gaussian function to estimate noise in that stack image.  

\begin{figure}
\begin{flushleft}
\rotatebox{0}{\includegraphics[width=0.5\textwidth]{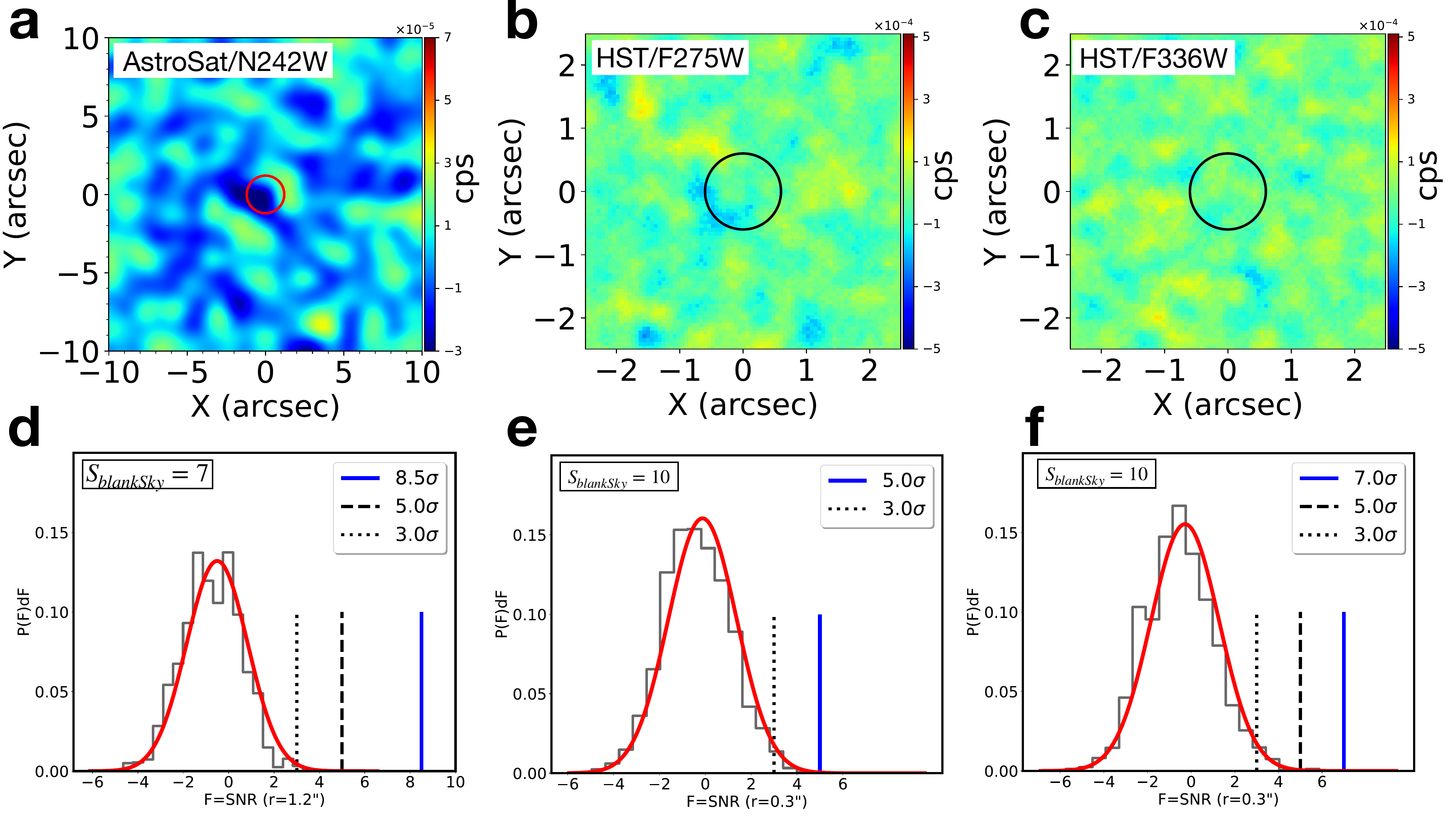}}
\caption{{\bf Blank-Sky stack:} 60 random locations, away from any source are selected from the UVUDF/F275W, F336W mosaics and AUDF-South N242W filter mosaics. From these 60 cutouts, 10 are chosen randomly to produce a stack in the HST filters. The same has been done for the N242W filters with randomly drawing 7 objects at a time. This has been repeated over 7000 times for N242W, 8000 for F275W and 7000 times for the F336W filter. Panel a, b, c refer to the mean stack image chosen randomly from the iteration. In panel a, the radius of the circle is 1.2" and on panels b, c, they are 0.6". Panels d, e and f display the probability density of the SNR estimated for the central aperture fluxes. Note that the integration of P(F)dF over the range of the histogram gives a value 1.}
\label{fig:blankSky}
\end{flushleft}
\end{figure}

The whole stacking process is repeated 7000 times for N242W, F336W filters and 8000 times for F275W filter to construct a robust distribution of central-aperture fluxes from the stack of these blank-sky cutouts. The resulting distribution is fitted with a Gaussian to find out how many events are there above $3\sigma$ or $5\sigma$. Figure~\ref{fig:blankSky} displays the distribution of central aperture fluxes for each filter. For each filter, we count the number of apertures that exceeds $SNR>3$. In N242W, we found 5 apertures with $SNR \ge 3$ (with P=0.0007) and zero apertures with $SNR \ge 5$ and above. In F336W filter, there are 163 apertures with $SNR \ge 3$ (corresponding probability P=0.023); 9 apertures with $SNR \ge 5$ (with P=0.0013) and zero apertures with $SNR \ge 7$. So a stacked detection at $3\sigma$ in F336W filter is likely to be contaminated but strong evidence for a stack with detection significance greater than $5\sigma$. Similarly,  we found 146 apertures with $SNR \ge 3$ (with P=0.018); 7 apertures with $SNR \ge 5$ (with P=0.00087). From about 7000 - 8000 such blank-sky realizations, we find that the probability of obtaining a $3\sigma$ signal by chance is $P\simeq 2\%$ in individual filters F275W and F336W, and $\simeq 0.07\%$ for N242W. None of the 7000 blank-sky stacks for N242W or F336W exceeded $5\sigma$, and only 7 out of 8000 blank-sky stacks for F275W reached $5\sigma$, corresponding to a probability $P<0.1\%$. Since UVIT and HST/UVIS are two different instruments with independent noise characteristics, the joint (or coincidence) probability of obtaining a signal $\ge 3\sigma$, is $P_{joint} = P_{N242W} \times P_{F336W} = 0.0007\times 0.023 = 1.6\times 10^{-5}$. In other words, the recovery of co-spatial signals at comparable significance (with $S/N>4.5$) across multiple filters makes it unlikely that the detection arises from noise fluctuations in any individual band, faint contaminants, or artifacts introduced by the stacking procedure, thereby strengthening the robustness of the result.

\subsection{Jacknife Resampling Test}

\begin{figure}
\begin{flushleft}
\rotatebox{0}{\includegraphics[width=0.5\textwidth]{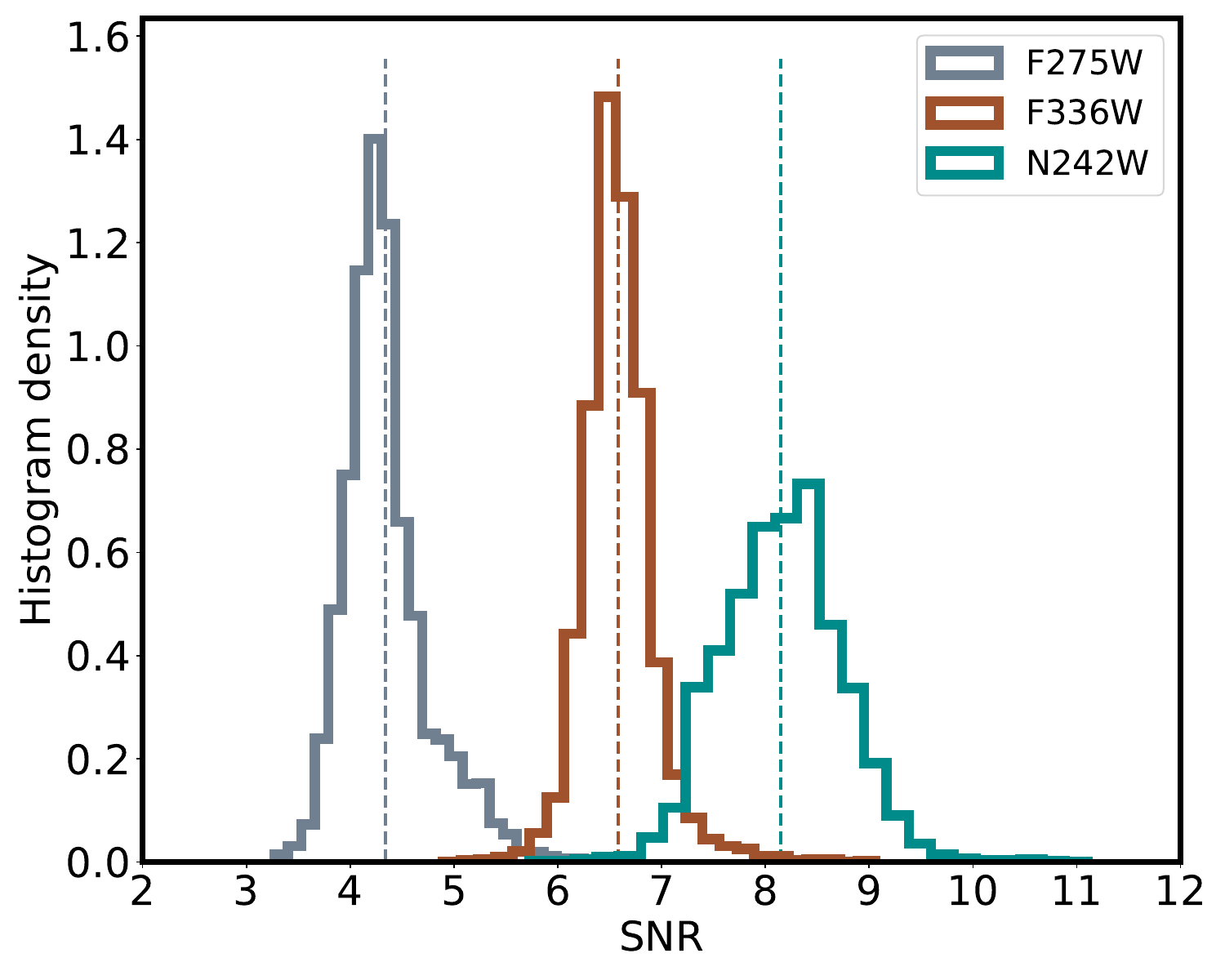}}
\caption{\textbf{Jacknife resampling:} performed for HST/F275W, F336W filters and AstroSat/UVIT/N242W filter. For the HST filters, the stack sample size is 10 from which one is dropped randomly and the rest of the sample galaxies are stacked. The same is carried out for the N242W filter for which the sample size is 7. The histogram shows distribution of the detection SNR estimated within an aperture of radius 0.3" for HST filters and 1.2" for N242W filter. The min (max) of the SNR in each case are 5.72 (11.1), 3.26 (6.5) and 4.9 (9.0) for N242W, F275W and F336W filters. The vertical dashed lines mark mean SNR.}
\label{fig:Onedrop}
\end{flushleft}
\end{figure}

In order to assess the influence of individual objects on the stacked signal, we perform a jackknife-style resampling analysis, in that we randomly exclude one galaxy at a time and stack the remaining objects. The same stacking procedure was applied (section~\ref{sec:stack}). Ideally, these processes would lead to a combination of $^{10}C_{9} = 10$ for HST Bronze sample and $^{7}C_{6} = 7$ for AstroSat Bronze sample. We repeat this 5000 times for each filter, each time with a different combination of galaxies. Each iteration leads to a different outcome due to changes in the background noise measurement. As explained above, during every iteration, we fill the masked regions (i.e., the segmented areas) with pixels drawing from neighbouring unmasked regions and this is done randomly. In other words, each iteration produces a different random realization of the stack background map. The background RMS is then estimated by fitting a Gaussian profile to the histogram of fluxes from randomly placed apertures of fixed size (size is the same within which we estimate the flux from the central aperture). The outcome of the whole iteration is shown as a histogram of the signal-to-noise ratio (SNR) for the central fluxes in three filters (N242W, F275W and F336W), see Figure~\ref{fig:Onedrop}. The mean (median) of the SNR distribution are 8.15 (8.16), 4.34 (4.28), and 6.58 (6.55) for N242W, F275W and F336W filters respectively. Given these numbers and full sample $SNR_{N}$ with N being the sample size, it is possible to estimate the Jackknife bias as:

\begin{equation}
    bias_{J} = \frac{N-1}{N} (\overline{SNR_J} - SNR_{N}),
\end{equation}
\noindent and the bias-corrected SNR as:

\begin{equation}
    SNR_{J} = SNR_{N} - bias_{J}.
\end{equation}

\noindent Then the bias and bias-corrected SNR are given by $bias_{J} =-0.30, -0.59$, and $-0.37$ and $SNR_{J} =8.8, 5.59$ and $7.37$ for N242W, F275W and F336W filters respectively. The negative bias indicates that the full-stack SNR may have been slightly underestimated. In all iterations, the resulting stacked flux remains above the $3\sigma$ level, demonstrating that no single object is responsible for driving the detection. This confirms that the stacked signal representing ionizing photons (in three filters) is a collective mean property of the sample rather than an artifact of a single outlier.

\section{IGM modelling: Lyman Valley transmission and Monte Carlo simulation}
\label{sec:IGM}

\subsection{Analytic model}
The Lyman valley refers to the broad depression in the extreme ultraviolet ($\lambda_{rest} < 912 \AA$) continuum of high-redshift sources, caused by the cumulative opacity of the intergalactic medium (IGM). First characterized in detail by \cite{Moller90}, the valley arises from the superposition of high-order Lyman series line blanketing and Lyman continuum (LyC) absorption by intervening H{\sc i} systems. The effective transmission through the intergalactic medium (IGM) along the line of sight to a source at redshift $z_{s}$ is given by:

\begin{equation}
    T(\lambda_{obs}, z_{s}) = e^{-\tau_{eff}(\lambda_{obs},z_{s})},
\end{equation} 

where $\tau_{eff}$ is the cumulative optical depth from discrete H{\sc i} absorbers (e.g., the Lyman forest, LLSs, and DLAs) along the line of sight to a source. The effective optical depth can be computed by adding contributions from both discrete Lyman series lines and Lyman Continuum. In other words,

\begin{equation}
    \tau_{eff}(\lambda, z_s) = \tau_{lines}(\lambda, z_s) + \tau_{c}(\lambda, z_s)
\end{equation}

Following \cite{Moller90}, we consider only the Lyman continuum absorption i.e., the opacity due to the photoelectic absorption by neutral hydrogen gas below the Lyman limit ($\lambda_{LL}=911.8$ \AA) and it is given by:

\begin{equation}
    \tau_{c}(\lambda, z_s)=\int_{0}^{z_s}{dz}\int_{N_{min}}^{N_{max}}{dN_{HI} f(N_{HI},z})[1 - e^{- N_{HI} \sigma_{HI}(\lambda) }],
    \label{eq:tau_c}
\end{equation}

\noindent where $\sigma_{HI}(\lambda)= \sigma_{0}\left(\frac{\lambda}{\lambda_{LL}}\right)^{3}$ refers to the hydrogen photoionization cross section for $\lambda < \lambda_{LL}$ and $\sigma_{0}=6.3\times10^{-18}$~cm$^{2}$. In the above equation, we consider a simple model for the absorber distribution: 

\begin{equation}
    f(N_{HI},z) = A\times(1+z)^{\gamma} \times N_{HI}^{-\delta},
    \label{eq:densityfunction}
\end{equation}

\noindent where $\delta \sim 1.3 - 2.0$ and $\gamma \sim 2 - 3 $ and $A$ is the normalization constant. Equation~\ref{eq:tau_c} can be evaluated numerically given a distribution of H{\sc i} absorbers. If we assume an average value of the H{\sc i} column density (instead of a power-law), the integral can be solved analytically, providing the necessary insight into the Lyman valley, in which the effective optical depth is directly proportional to $\left(\lambda/\lambda_{LL}\right)^3$, see \cite{Moller90}. In Figure~\ref{fig:LVtransmit}, we show numerically integrated transmission spectrum for z=5.94 assuming model parameters for the absorber distribution as $A = 4 - 8 \times 10^6$, $\gamma= 2.3$, $\delta = 1.5$ and column density varying from $\log N_{HI}=11 - 20$ (assuming no distinction between different types of absorbers, for simplicity of analytical calculation). The chosen range of normalization parameter ($A$) incorporates the observed number density of HI clouds at $z\sim 6$ that shows $\sim 0.5$-dex variation \citep[Figure 2 in][]{Inoueetal2014}. Overall, the absorber densities produced by these model parameters are somewhat close to those in Inoue et al. 2014 \citep[see for details Figure 1 in][although they are not well constrained at this redshift]{Inoueetal2014}. 
It can be seen from the figure that even without the line blanketing from the Lyman forest, the IGM continuum opacity alone can produce a significant suppression of extreme-UV flux below the Lyman limit. However, the transmission starts rising at shorter wavelength (below $500 \AA$ in our case). 

\subsection{Monte Carlo simulation}
The transmission of the IGM was modeled based on an empirical number density function of H~{\sc i} absorbers \cite[see equation 4 in][]{Inoueetal2014}, composed of three populations of H~{\sc i} absorbers: Ly$\alpha$ forest (LAF), Lyman limit systems (LLs) and damped Ly$\alpha$ absorbers (DLA). Following \cite{Inoueetal2014}, the distribution of absorbers are modelled with a sum of two different power-laws - first component refers to the "LAF" and the second one "DLA". In both cases, the H~{\sc i} column density covers a wide range from $12 < \log N_{HI} < 23$ \citep[readers are referred to][for further details]{Inoueetal2014}. The mean transmission model based on the number density function reproduces the observed mean transmission as a function of redshift very well. For a source at $z=5.94$, the predicted mean transmission (averaged over 10000 sightlines) in the N242W filter is virtually zero. The detection in N242W, F275W and F336W filters suggests a large deviation from the mean. We hence performed a Monte Carlo simulation \citep{Inoue08} to examine the probability to have a high transmission. These MC simulations assume a completely random distribution of H~{\sc i} absorbers along a given line-of-sight. However, the real absorbers are clustered with respect to each other and correlated or anti-correlated with the source itself in a scale close to the source. These effects would enhance the transmission fluctuations with respect to the uniform random case (considered here). 
Therefore, the obtained probability of a high transmission is a lower bound. This clustering is expected to manifest as correlations in the IGM optical depth, and has been identified on scales of at least $\sim$100 cMpc in recent work \citep{Meyeretal2025}. It may help explain the unexpectedly high transmission, and consequently low Ly$\alpha$ optical depths, observed along some quasar sightlines at $z \gtrsim 5.9$ based on the XQR30 survey \citep{Bosmanetal2022, DOdoricoetal2023}.

After 10000 Monte Carlo trials, we found the minimum IGM attenuation through the N242W filter to be 4.5 mag, which is too high attenuation (low transmission) to detect any flux in N242W. If we neglect the contribution of the "DLA" component, the minimum IGM attenuation in N242W becomes 1.4 mag (see Figure~\ref{fig:LVtransmit}). Although the probability of this case is extremely low at face value in our simulation ($p<10^{-4}$), this could be due to our assumption of a random distribution of absorbers. 
In the $z\sim 6$ Universe, there is a strong scatter in the opacity along different quasar sightlines, indicating a non-uniform distribution of absorbers, and the clustering of absorbers is expected to drive the scatter from different line-of-sight attenuations \citep{Bosmanetal2022}. A low-opacity line of sight is extremely rare in current IGM models (1/10000) but it may be higher in more complex models including clustering (not considered in standard IGM models).
Indeed, in the real Universe, the IGM transmission should fluctuate much more \citep{Crightonetal2015} than the uniform random distribution assumed in our simulation - an implication from the stacking detection in three filters at $z=5.94$. Furthermore, these Monte Carlo simulations only include H~{\sc i} absorbers, the contribution from Helium absorbers is not implemented. If Helium on the line-of-sight is neutral, i.e. He~{\sc i}, the transmission at rest-frame wavelengths shorter than 584~\AA\ should be reduced by He~{\sc i} "Ly$\alpha$" absorption. This would significantly lower the transmission in N242W, that is inconsistent with the detection in that band. Therefore, the N242W, F275W and F336W stacking detections together suggest that Helium is not neutral along the line-of-sight.  


\begin{table*}
\centering
\setlength{\tabcolsep}{4pt}
\resizebox{\textwidth}{!}{
\begin{tabular}{cccccccccc}
\hline
Label & $M_{\mathrm{UV}}$ & $\beta$ & $L_{\mathrm{H}\alpha}$ &
$\log_{10}\xi_{\mathrm{ion}}^{0}$ &
$\log_{10}\xi_{\mathrm{ion}}^{\mathrm{true}}$ &
EW(H$\beta$) & EW([O III]) &
$12+\log_{10}(O/H)$ & $12+\log_{10}(O/H)$ \\
& & & ($10^{41}$ erg s$^{-1}$) &
(Hz erg$^{-1}$) &
(Hz erg$^{-1}$) &
(\AA) & (\AA) &
($\widehat{RNe}$) & ($\widehat{R}$) \\
\hline
G1  & $-19.21 \pm 0.04$ & $-2.35 \pm 0.29$ & $5.47\pm0.28$ & $25.28\pm0.03$ & $25.76\pm0.02$ & $426.7 \pm 84.7$ & $4631.4\pm 470.8$ & -- & -- \\
G2  & $-17.66 \pm 0.12$ & $-3.83 \pm 1.03$ & $0.31\pm0.07$ & $24.64\pm0.11$ & $26.2\pm0.05$ & $178.9 \pm 59.4$ & $1128.8 \pm 215.4$ & -- & -- \\
G3  & $-20.33 \pm 0.04$ & $-1.92 \pm 0.04$ & $4.01 \pm 0.27$ & $24.7\pm0.03$ & $25.27\pm0.01$ & $110.25 \pm 7.8$ & $994.0 \pm 65.8$ & $7.10\pm0.26$ & $7.71\pm0.33$ \\
G4  & $-18.48 \pm 0.35$ & $-3.17 \pm 0.37$ & $0.50 \pm 0.06$ & $24.53\pm0.15$ & $25.91\pm0.14$ & $<37.2$ & $383.9 \pm25.3$ & -- & -- \\
G10 & $-19.55 \pm 0.03$ & $-3.75 \pm 0.73$ & $1.85\pm0.05$ & $24.67\pm0.02$ & $25.53\pm0.02$ & $212.6 \pm 14.9$ & $1439.5\pm46.2$ & $7.38 \pm0.24$ & $7.49\pm0.28$ \\
\hline
\end{tabular}}
\caption{\textbf{Physical properties of the Gold sample:} The equivalent widths (EW) are estimated in the rest frame (units of \AA). Details about the oxygen abundance measurement can be found in Section~\ref{sec:abundance}. $\xi_{\mathrm{ion}}^{\mathrm{true}}$ is defined in Section~\ref{sec:fesc}.}
\label{tab:physical_properties}
\end{table*}

\begin{figure}
\begin{flushleft}
\rotatebox{0}{\includegraphics[width=0.5\textwidth]{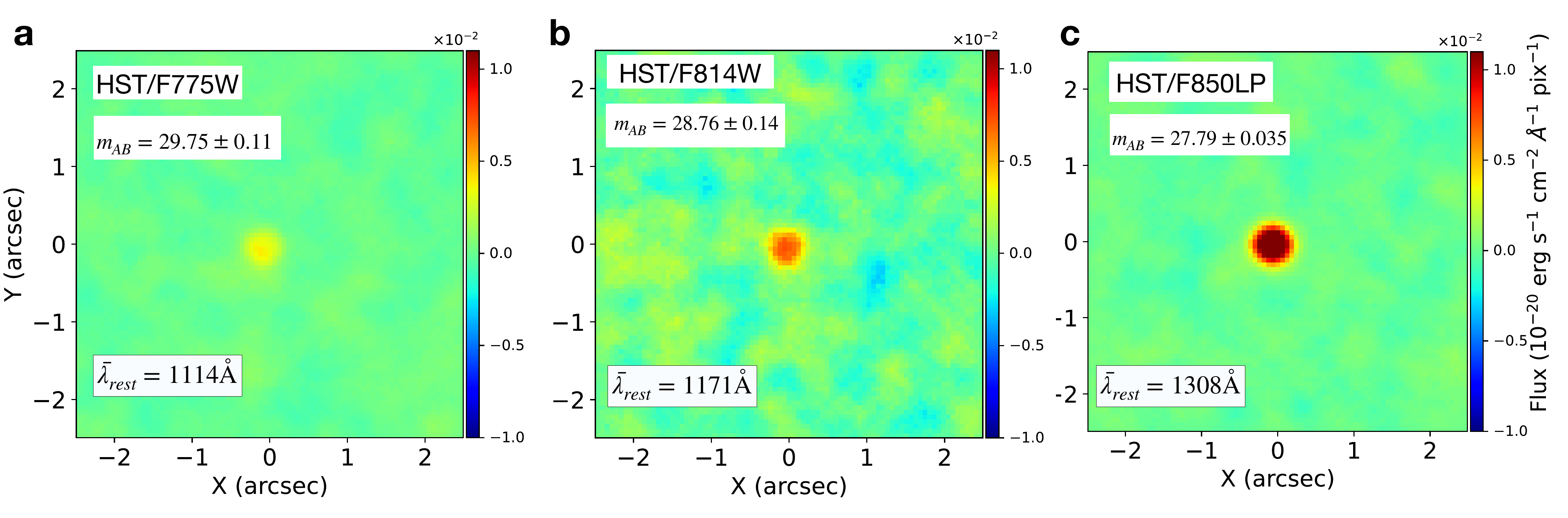}}
\caption{\textbf{Non-ionizing UV continuum:} Stack in ACS/WFC/F775W, F814W and F850LP filters corresponding to the rest-frame wavelength of $\bar{\lambda}_{rest}=1114\AA, 1171\AA$ and $1308 \AA$ with SNR=$10.4\sigma$, $7.8\sigma$ and $30.8\sigma$ respectively (see Table~\ref{tab:photometry_stack}).}
\label{fig:UVcontinuum}
\end{flushleft}
\end{figure}

\section{Physical properties of the galaxies in the stack}
\label{sec:physical}

\subsection{UV continuum and $\beta$ slope}
\label{sec:betaslope}

To characterize the non-ionizing UV continuum and probe IGM attenuation blueward of Ly$\rm \alpha$, we stacked deep HST/ACS (F775W, F814W, F850LP) images with MUSE Ly$\rm \alpha$ narrow-band data for the Bronze sample. For the MUSE LAEs \citep{Baconetal2023}, the median Ly$\rm \alpha$ luminosity is $L_{\mathrm{Ly}\alpha} = (6.8 \pm 0.36) \times 10^{41}$~erg~s$^{-1}$, with a median (mean) equivalent width of $83.0 \pm 40.9$~\AA\ ($104.2 \pm 32.6$~\AA). The Ly$\rm \alpha$ narrow-band stack yields a strong $22\sigma$ detection, and the rest-frame UV continuum is clearly detected between $1114 - 1308~\AA$ (see Figure~\ref{fig:Stack_Lya}, Figure~\ref{fig:UVcontinuum}; and Table~\ref{tab:photometry_stack}). Photometry, sampling both sides of Ly$\rm \alpha$ emission, reveals strong attenuation in the bluer filters: F814W is $\sim$1 mag fainter, and F775W $\sim$2 mag fainter than F850LP. Assuming a flat UV continuum, we estimate effective optical depths of $\tau_{\rm eff} \approx 0.91$ and $1.84$, respectively. The detection of residual flux shortward of Ly$\rm \alpha$ indicates that the Gunn-Peterson trough \citep{Gunn65} is not fully saturated at this redshift. This is consistent with a picture of inhomogeneous reionization, in which bubble-like reionization topology \citep{Leeetal2008} allow transmission along some sightlines despite a substantial neutral hydrogen fraction in the cosmic average of this epoch \citep{Meyeretal2025}.

The rest-frame UV ($1500,\AA$) luminosity of the HST Bronze stack is $M_{\rm UV} = -18.77\pm0.05$. This corresponds to a low-luminosity population ($\sim0.1L_{*}$), $\sim$2.1 mag fainter than $M_{\rm UV}^{*}$ at $z\sim6$ \citep{Bouwens21}. The non-ionizing UV continuum slope measured from the stacked Bronze-sample photometry in the F850LP, F105W, F125W, and F140W filters is $\beta = -2.5 \pm 0.18$. In terms of UV magnitude and $\beta$-slope, our Bronze stack outcome closely matches with the compact star-forming galaxies that are blue ($\beta \sim -2.17$) and faint ($M_{\rm UV} = -18$) in the redshift range $6 < z <9$ \citep{Mascia24}. In contrast, the Gold-sample galaxies exhibit significantly bluer UV continua, with a median $\beta = -3.17 \pm 0.38$. Of these, G2, G4 and G10 have extreme bluer slopes, $\beta \leq -3$, (see Table~\ref{tab:physical_properties}). Such extreme UV slopes are indicative of very young, low-metallicity, and nearly dust-free stellar populations, with minimal contribution from nebular continuum emission \citep{Schaerer2003,Bouwensetal2010}. 

\begin{figure}
\rotatebox{0}{\includegraphics[width=0.52\textwidth]{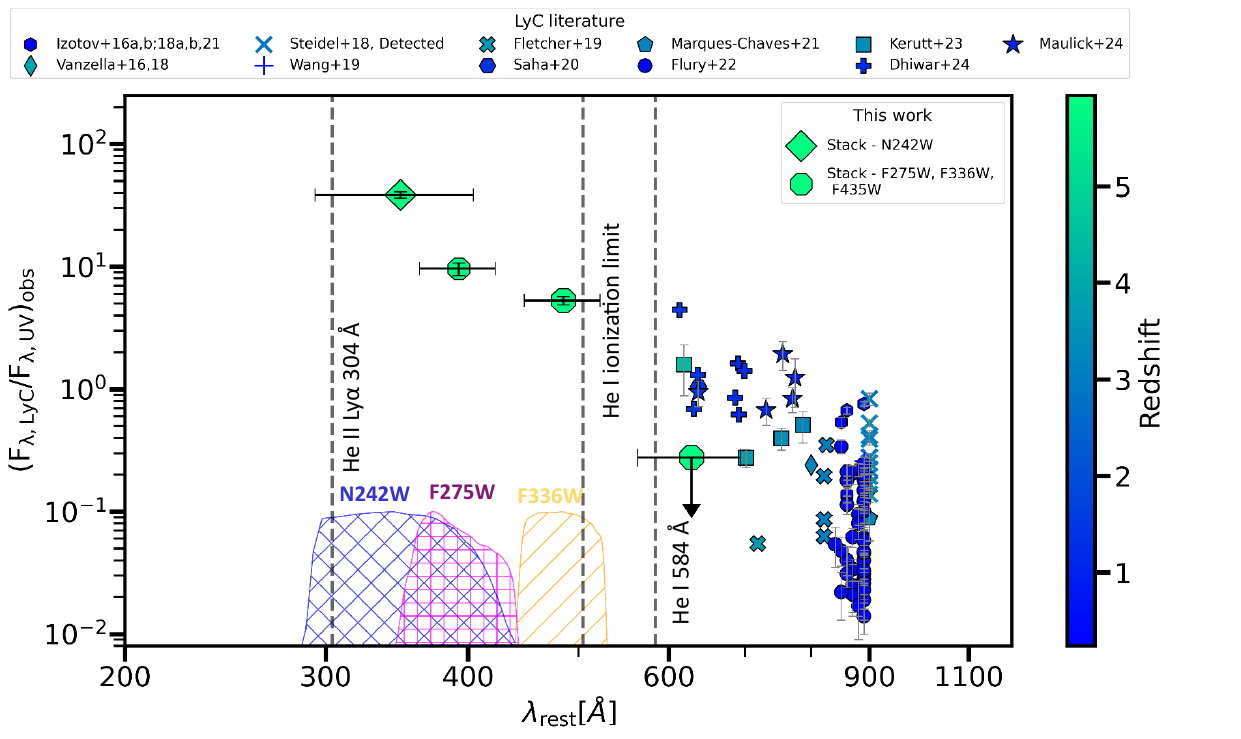}}
\caption{{LyC to UV continuum (at $1500 \AA$) flux density ratio} (as observed) as a function of rest-frame wavelength - {showing the demographics of all the detected ionizing photons from literature \citep{Vanzella16,Vanzellaetal2018,Izotov2016,Izotov16b,Izotov18,Izotov18b,Izotov21,Steideletal2018,Wang19, Fletcheretal2019,Sahaetal2020,MarquesChaves21,Keruttetal2023,Dhiwar24, Maulick25}}. Each data point on this plot is color coded according to its redshift. {For consistency with the stacked ratios in this work, the literature points are not corrected for IGM attenuation.} The horizontal bars on the data points represent the FWHM of the corresponding filter. All three filters are blue-shifted according to the mean redshift z=5.94.}
\label{fig:Lycratio}
\end{figure}

\begin{figure}[h]
\begin{flushleft}
\rotatebox{0}{\includegraphics[width=0.5\textwidth]{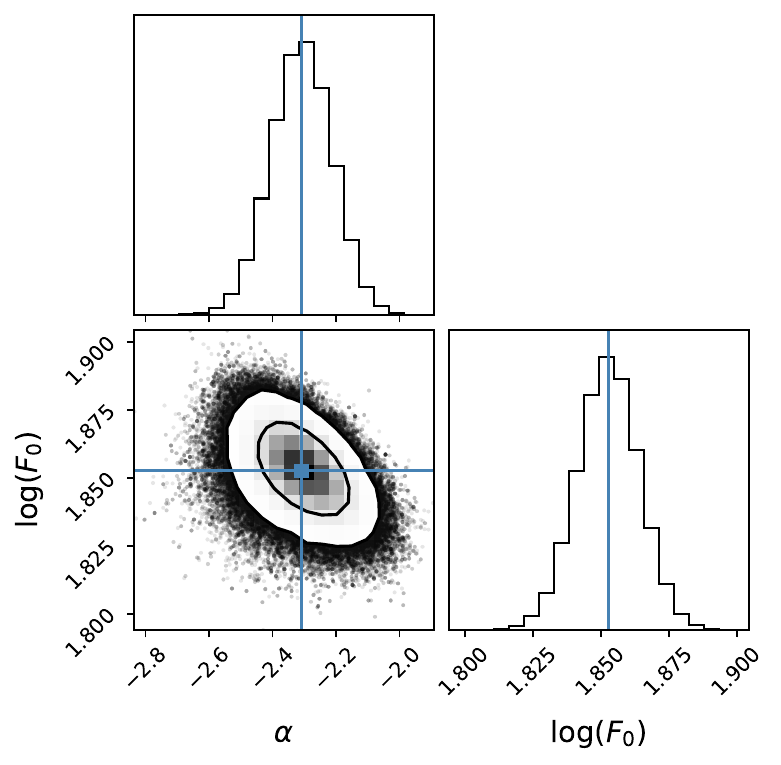}}
\caption{{\bf Spectral Slope:} An MCMC analysis for estimating the ionizing spectral slope. }
\label{fig:slope}
\end{flushleft}
\end{figure}

\subsection{LyC-to-UV ratio and Slope of the ionizing spectrum}
\label{sec:MCMCfit}

The observed flux-density ratio of ionizing to non-ionizing radiation, $(F_{\lambda, \mathrm{LyC}} / F_{\lambda, \mathrm{UV}})_{\rm obs}$, serves as a key empirical proxy for the escape of LyC photons \citep{Steidel01}. As shown in Figure~\ref{fig:Lycratio}, this ratio increases progressively at shorter wavelengths probed by the F336W, F275W, and N242W filters. Various surveys of LyC emitters probe different portions of the rest-frame ionizing spectrum depending on filter bandpasses and detector configurations. Most of these earlier studies sample wavelengths $\lambda_{rest} > 500~\AA$, while our detections uniquely extend to shorter wavelengths ($\lambda_{rest} < 500~\AA$), providing access to a previously unexplored regime of the ionizing continuum. Our stacking results demonstrate, for the first time, that ionizing photons at rest-frame wavelengths below $500~\AA$ are not only produced in appreciable quantities but may also escape from galaxies during this epoch — marking the Cosmic Morning, a transitional phase between the emergence of galaxies and the end of reionization. 

\begin{figure*}
\begin{flushleft}
\rotatebox{0}{\includegraphics[width=0.7\textwidth]{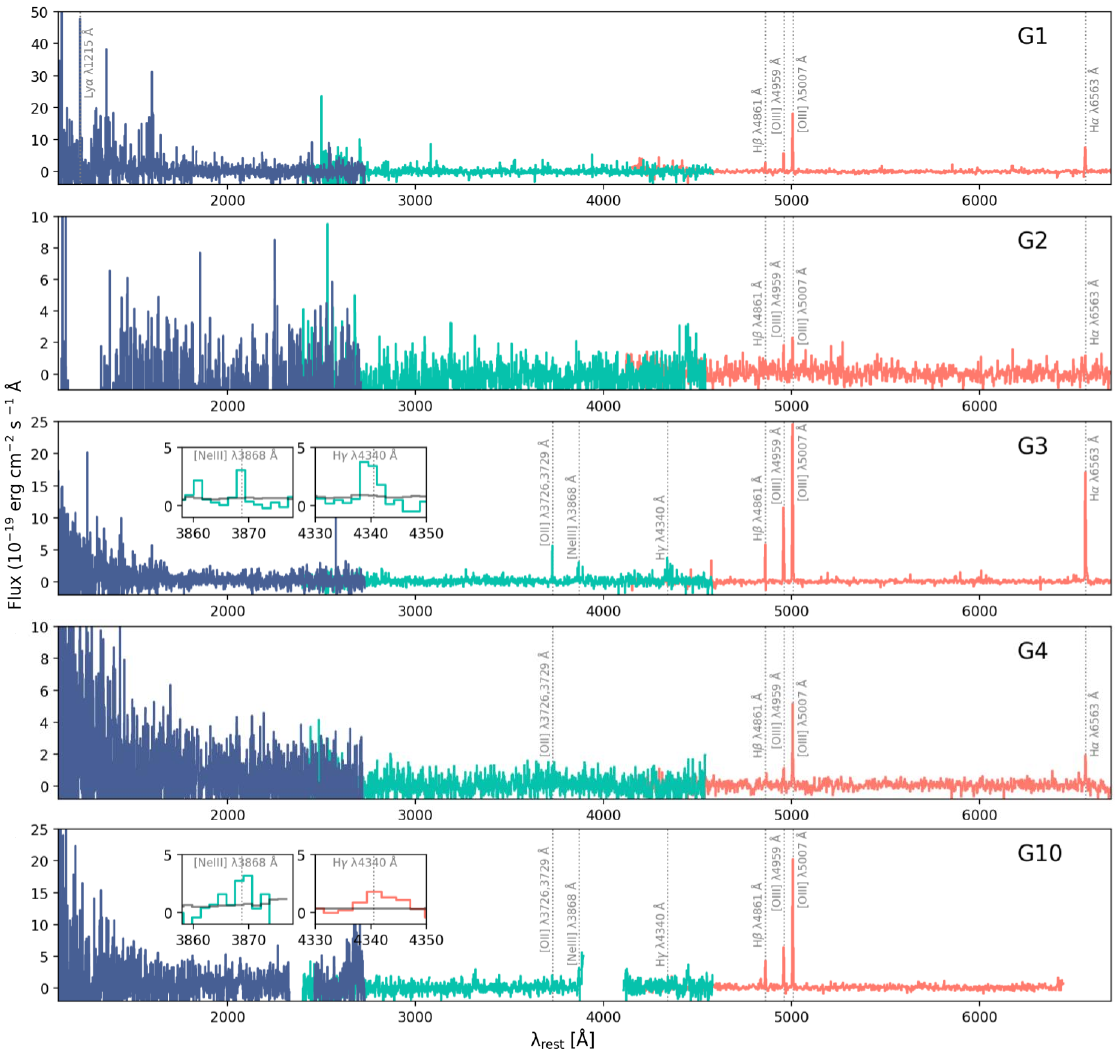}}
\caption{{\bf Emission-line spectra:} 1D JWST/NIRSPEC grating spectra (G140M/F070LP: blue; G235M/F170LP: green; G395M/F290LP: red) for the five galaxies in our gold sample. The emission-lines detected with S/N$>$3 are marked in grey. For G3 and G10, the insets show the zoom-in of the spectra around the [NeIII]$\lambda$3686 \AA~and H$\rm\gamma$ lines, along with their corresponding uncertainty (grey).}
\label{fig:R1000_spectra}
\end{flushleft}
\end{figure*}

Our stack detections represent one of the rare cases in which ionizing flux is robustly measured across three filters. To the best of our knowledge, there are no previous reports of such detections obtained simultaneously in even two independent filters. In other words, these observations can, in principle, be used to determine the slope of the ionizing spectrum. However, unfolding the intrinsic ionizing spectrum requires modelling the combined effects of the ISM, IGM transmission, and escape fraction. Under a simplified assumption of negligible dust ($\beta = -2.5$), unity escape fraction, and IGM transmission as modelled in Figure~\ref{fig:LVtransmit}, we fit the observed ionizing fluxes using an MCMC analysis (described below). 

We utilize the most favourable IGM transmission model (see Figure~\ref{fig:LVtransmit}) to obtain the intrinsic flux value at the three filters (N242W, F275W and F336W) where we detect ionizing flux and at F435W filter where have an upper limit using the following equation:

\begin{equation}
    F_{obs}(\lambda) = F_{int}(\lambda) \times T_{IGM}(\lambda) 
\end{equation}

Here, we model a relation between the ionizing wavelength and normalized flux (normalized by the measured flux value at rest-frame 1500 \AA, denoted as $F_{1500}$) using a linear fit in the log-log space incorporating three detections and an upper limit.

\begin{equation}
    \frac{F_{int}(\lambda)}{F_{1500}} = F_{0} \left(\frac{\lambda}{\bar{\lambda}}\right)^{\alpha_{\lambda}},
\end{equation}

\noindent then $\log(F_{0})$ becomes the intercept and $\alpha_{\lambda}$ is the spectral slope. Here $\bar{\lambda}=463.6 \AA$ corresponds to the arithmetic mean of rest-frame wavelength corresponding to the four data points. The fit is performed using a Markov Chain Monte Carlo (MCMC) sampler implemented in the "emcee" package. Measurement uncertainties along both axes were included in a Gaussian likelihood, and the upper limit was treated as a censored data point using a one-sided Gaussian likelihood. Note that for uncertainties along the wavelength axis, we used the filter FWHM. The posterior probability distribution includes the slope ($\alpha_{\lambda}$), intercept ($\log F_{0}$), and true $\lambda/\bar{\lambda}$ - values for each point. We impose uniform priors for the slope and intercept: $-4 < \alpha_{\lambda} < 4$ and $-3<\log F_{0} < 3$. The MCMC sampling was run with 64 walkers for 10000 steps; the first 1000 steps of each chain were discarded (warm-up phase of MCMC chain). The resulting posterior distributions is visualized using a corner plot, which displays the marginalized one-dimensional distributions and the two-dimensional joint confidence contours. These contours correspond to the 68\% and 95\% confidence regions and illustrate the covariance between slope and intercept (Figure~\ref{fig:slope}). The values of the slope and intercept are given by: ($\alpha_{\lambda}, \log F_{0}$)= ($-2.3\pm0.1, 1.85\pm{0.01}$). The resulting ionizing spectral slope $\alpha_{\lambda} = -2.3 \pm 0.1$, is significantly steeper than the spectral indices typically observed for AGN in the rest-frame EUV band \citep{Stevansetal2014}. In order to get some insight, we consider a hot blackbody source \citep{Olivieretal2022} with temperature T for shorter wavelength regime:

\begin{equation}
    I_{B}(\lambda) = \frac{2 hc^2}{\lambda^{5
    }} e^{-hc/{\lambda k_{B}T}}
\end{equation}

\noindent  Using the log derivative, we get a spectral slope for a hypothetical source at T=T$_{s}$ and $\lambda=\lambda_{s}$

\begin{equation}
    \frac{d\log I_{B}}{d\log \lambda} = -5 + \frac{hc}{\lambda_{s} k_{B} T_{s}}
\end{equation}

Equating this to our slope $\alpha_{\lambda}$, we get 
\begin{equation}
    T_{s} = \frac{hc}{(5+\alpha_{\lambda})k_{B}\lambda_{s}},
\end{equation}

\noindent substituting $\alpha_{\lambda}=-2.3$ and $\lambda_{s}=485$ \AA, we get $T_s = 1.09\times 10^{5}$K. This might indicate the effective temperature of the radiation field that ionized the gas. 

\begin{figure*}
\rotatebox{0}{\includegraphics[width=0.9\textwidth]{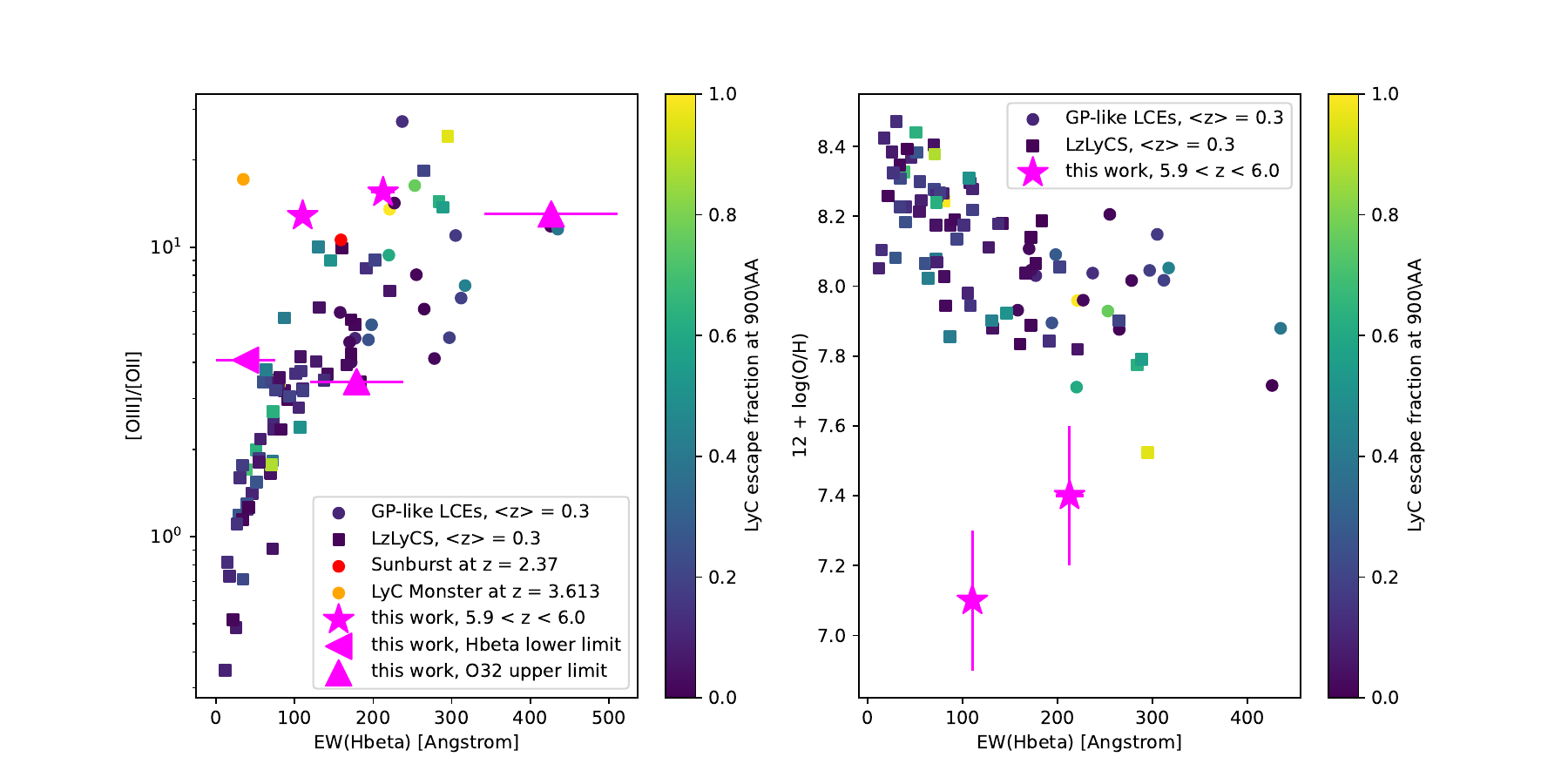}}
\caption{{\bf Comparison with Low-z leakers}: At $\langle z \rangle\sim 0.3$, \citep[LzLCS+ sample,][]{Izotov2016,Izotov16b,Izotov18,Wang19,Izotov21,Fluryetal2022a}, Sunburst Arc at $z=2.37$ \citep{RiveraThorsenetal2019}, and the so-called LyC Monster \citep{MarquesChavesetal2022} at $z=3.61$. \textbf{Left panel}: O32 vs EW H$\beta$. \textbf{Right panel}: gas-phase metallicity vs EW H$\beta$.}
\label{fig:ComparisonPlot}
\end{figure*}

\subsection{Chemical abundance and age measurement}
\label{sec:abundance}
For the five galaxies in our gold sample, we utilised their JWST/NIRSpec medium resolution grating spectra (G140M/F070LP, G235M/F170LP: and G395M/F290LP), obtained as part of the JADES survey \citep{Bunker23,Curtis-Lake25} and publicly available via the Mikulski Archive for Space Telescopes (MAST), for flux measurements of emission-lines of interest, and subsequent abundance determination where possible. Each 1D JWST/NIRSpec spectrum is delivered by MAST as a Level 3 calibrated data product, following an optimal extraction procedure from the calibrated 2D spectra. The spectra are flux calibrated with a $\sim25$\% absolute flux uncertainty with the emission-lines in the grating spectra having mean flux lower by $\sim8$\% compared to the prism spectra \citep{Scholtz25}. 

The 1D spectra for the five galaxies are shown in Figure~\ref{fig:R1000_spectra}. We utilise the standard specutils package \citep{specutils19} to determine emission-line fluxes and SNR. The measured line fluxes for all lines from the 1D spectra from MAST for all five galaxies are consistent with those from the JADES DR4 data release \citep{Scholtz25} with 3-pixel extraction.  The noise is highest for the G140M/F070LP spectra that had lowest exposure times in the survey, and lowest for the G395M/F290LP spectra with the highest exposure times. The  [OIII]$\lambda\lambda$4959,5007 \AA~ doublet is clearly detected for all five galaxies while the H$\alpha$~line is clearly detected for all but one galaxy (it lies in a chip-gap for G10; we utilise the prism spectrum of G10 shown in Figure~\ref{fig:MUSEJWST} for the H$\alpha$~line flux measurement). The measured FWM of the 
${[\mathrm{O\,III}]\,\lambda5007}$ ranges from $192 - 372\AA$ in the Gold sample galaxies, while for ${[\mathrm{H\,\alpha}]\,\lambda6563}$, they are $176 - 337\AA$ showing no signs of broadening. These galaxies also lack the high-ionization UV lines such as ${[\mathrm{C\,IV}]\,\lambda1550}$, ${[\mathrm{N\,V}]\,\lambda1250}$ that are typical AGN-like \citep{Feltre17} in the combined MUSE and JWST medium resolution grating spectra.
The [OII]$\lambda\lambda$3727,3729 \AA~lines have S/N$>$3 for G3, G4 \& G10, while the [NeIII]$\lambda$3686 \AA~and H$\rm\gamma$ lines are additionally detected with S/N$>$3 for G3 \& G10. Using the H$\beta$ and H$\rm\gamma$ lines for estimating the nebular dust attenuation, we find G3 \& G10 are both consistent with negligible dust attenuation.  

While no auroral lines are detected, the [OII]$\lambda\lambda$3727,3729 \AA, [NeIII]$\lambda$3686 \AA~and H$\rm\gamma$ line flux measurements allow for strong-line estimates of the O abundances of G3 \& G10 via the $\widehat{RNe}$ strong-line calibration \citep{Scholte25}. This calibration has been shown to allow robust estimates of O abundances for both low-z and high-z (out to z$\sim8$) star-forming galaxies by virtue of being less affected by systematic evolution in ionization conditions with redshift as well as being less affected by varying attenuation laws. For G3, we estimate $12+\log{O/H} = 7.1\pm0.26$ ($\sim2.5$\% Solar) while for the G10, it is $12+\log{O/H} = 7.38\pm0.26$ ($\sim5$\% Solar). The uncertainty in the estimated abundance considers both the errors in the line flux measurements for these galaxies and the dispersion in the calibration \citep{Scholte25}, added in quadrature. We note that the determined abundances are robust against the error in the absolute flux calibration as the line flux ratios remain unaffected. The [OII]$\lambda\lambda$3727,3729 \AA, [OIII]$\lambda\lambda$5007,4959 \AA~and H$\rm\beta$ line flux measurements additionally allow strong-line O abundance estimates for the same two galaxies via the $\widehat{R}$ strong-line calibration \citep{Laseter24}, recalibrated for metal-poor (12+log(O/H)<8.2) galaxies by \cite{Scholte25}. For G3, we estimate $12+\log{O/H} = 7.71\pm0.33$ ($\sim10$\% Solar) while for the G10, it is $12+\log{O/H} = 7.49\pm0.26$ ($\sim6$\% Solar). The $\widehat{R}$ O abundance estimates are thus consistent with that of $\widehat{RNe}$ for G10, but slightly higher for G3. Nevertheless, both galaxies are very metal-poor in any case (2--10\% Solar).

The age of the stellar population is estimated using the equivalent width (EW) of the H$\beta$ emission line. To compute the EWs, we combine the H$\beta$~4861\AA\ line fluxes measured from the NIRSpec medium-resolution grating spectra with continuum estimates derived from the NIRCam photometric measurements, as the continuum is only weakly detected in the spectra. The H$\beta$~4861\AA\ line is redshifted to the middle of JWST/NIRCam F335M filter. Since F335M filter covers the bluer half of the F356W filter, we utilize these two filter combination directly to estimate the stellar continuum, assuming the continuum to be nearly constant within both filters.
Note that we do not apply dust corrections to the line fluxes. For G3 and G10, the observed H$\gamma$/H$\beta$ ratios indicate negligible dust. For the remaining galaxies in the Gold sample, where H$\gamma$ measurements are unavailable, we assume zero dust based on their blue UV continuum slopes \citep{Reddy18} and the fact that the upper limits on the H$\gamma$ line are consistent with a lower limit of negligible dust. For example, with $\beta = -2.5$ for the Bronze stack gives $E(B-V) \sim 0.02$ and for the Gold stack with $\beta = -3.17$,  $E(B-V) < 0$. Utilising H$\alpha$/H$\beta$ ratios for dust correction of extreme emission-line galaxies, including Lyman-$\alpha$ emitters at low-z, can cause an over-correction of the observed spectra in cases where Case B recombination does not hold \citep{Scarlata24}. We thus utilise the less affected H$\gamma$/H$\beta$ ratios. A comparison of line measurements from our galaxies to lower-redshift leakers is presented in Figure~\ref{fig:ComparisonPlot}: our galaxies share strong line properties of low-z LyC leakers, but have lower metallicities, as expected for these less evolved systems.

\section{Escape fraction and ionizing photon production efficiency}
\label{sec:fesc}

In section~\ref{sec:MCMCfit}, we modelled the observed ionizing spectrum using three measurements of stack at N242W, F275W and F336W corrected by the IGM transmission. The ionizing spectrum is given by

\begin{equation}
    F_{int} (\lambda) = F_{ion}^0 \left(\frac{\lambda}{\lambda_{0}}\right)^{\alpha_{\lambda}},
\end{equation}

\noindent where $\alpha_{\lambda} = -2.3\pm 0.1$; $\lambda_{0}=463.6\AA$ and $F_{ion}^{0} = F_{0}\times F_{1500} =4.67\times10^{-19}$~erg~s$^{-1}$~{cm}$^{-2}$~{\AA}$^{-1}$. Since the ionizing photon spectrum contains photons greater than 24.58 eV, and considering the LAEs in our sample are metal-poor, a fraction of these ionizing photons will be absorbed by H and He in the galaxy (assuming negligible dust). This depends on the number density and cross-section of the H and He. At a given energy (E > 13.6), we can define a factor that will take part in the H-atom ionization as:

\begin{equation}
    f_{H}(E) = \frac{n_{H} \sigma_{H} (E)}{n_{H} \sigma_{H} (E) + n_{He} \sigma_{He} (E) + n_{Z} \sigma_{Z} (E)},
\end{equation}

\noindent where the approximate relations for the photoionization cross sections for H and He are given by: 

\begin{equation}
    \sigma_{H}(E) = 6.3\times 10^{-18} (13.6/E)^3;\text{and} \ \sigma_{He}(E) = 7.8\times 10^{-18} (24.6/E)^3
\end{equation}

\noindent Since there are about 1 He atom in every 12 H-atom, $\frac{n_{He}}{n_{H}}=0.083$, which in turn results in a fraction of the H-ionizing photons as $f_{H}(E) = 0.62$ at the mean energy of the ionizing spectrum. Considering most of the galaxies in our sample are metal-poor (see section above), the metal-contribution in the denominator can be neglected. Then the total H-ionizing photons per second that escaped the galaxy can be obtained by solving the following integral:

\begin{equation}
\dot N_{\rm ion,H}^{esc} = \int_{\lambda_{\rm min}}^{\lambda_{\rm max}} \frac{\lambda\, F_{int}(\lambda)}{hc} \, f_H(\lambda) \, d\lambda,
\end{equation}

\noindent where $f_{H}(E) = f_{H}(hc/\lambda)$, with $hc = 1.986\times 10^{-8}$~erg~cm. $\lambda_{min}=288\ \AA$ and $\lambda_{max}=534\ \AA$ corresponding to the blue-end of N242W and red-end of F336W filters respectively. Then we can split the integral for $\dot N_{\rm ion,H}^{esc}$ using a step-function approximation for $f_H(\lambda)$ (since $\sigma_{He}(E<24.6~{\rm eV})=0$):

\begin{equation}
f_H(\lambda) =
\begin{cases}
1, & 912~{\rm \AA} \ge \lambda > 504~{\rm \AA} \quad (13.6 \le E < 24.6~{\rm eV})\\[2mm]
0.62, & 504~{\rm \AA} \ge \lambda > 228~{\rm \AA} \quad (24.6 \le E < 54.4~{\rm eV})
\end{cases}
\end{equation}

Then the total escaping H-ionizing photon rate becomes two separate integrals:

\begin{equation}
\dot N_{\rm ion,H}^{esc} = \frac{F_{ion}^{0}}{hc} \left[ 
\int_{504}^{534} \left( \frac{\lambda}{\lambda_0} \right)^{\alpha_\lambda} \lambda \, d\lambda 
+ 0.62 \int_{288}^{504} \left( \frac{\lambda}{\lambda_0} \right)^{\alpha_\lambda} \lambda \, d\lambda 
\right]
\label{eq:N_esc}
\end{equation}

Solving the integrals in Eq.~\ref{eq:N_esc}, we obtain $\dot N_{\rm ion,H}^{esc} = 2.14\pm0.04\times 10^{-6}$~photon~s$^{-1}$~{cm}$^{-2}$, where the error on the photon flux arises due to error on the slope of the ionizing photon spectrum. In other words, the net rate of ionizing photons escaped from the source is $\dot N_{\rm ion,H}^{esc} = 4\pi d_{L}^2 \times (2.14\pm0.04)\times 10^{-6}=8.3\pm0.15\times 10^{53}$~photon~s$^{-1}$, where the luminosity distance $d_L=1.76\times10^{29}$ cm based on the standard 737 cosmology. 

\noindent For the Gold sample, we measure the H$\alpha$ emission line fluxes from the JWST/NIRSpec to compute the number of ionizing photons that are absorbed in the recombination process as:

\begin{equation}
    Q_{rec}(H^{0}) = 7.28 \times 10^{11} L_{H\alpha},
\end{equation}

\noindent The $H\alpha$ luminosity is estimated using a simple mean of the measured fluxes as $F_{H\alpha} =62.7\pm 2.1 \times 10^{-20}$~erg~s$^{-1}$~cm$^{-2}$ corresponding to $L_{H\alpha} = 2.44\pm 0.082 \times 10^{41}$~erg~s$^{-1}$. This implies $Q_{rec}(H^{0}) = 1.78 \pm 0.06\times 10^{53}$~s$^{-1}$. Then the escape fraction can be estimated using the following relation:

\begin{equation}
    f_{esc} = \frac{\dot N_{\rm ion,H}^{esc}}{\dot N_{\rm ion,H}^{esc} + Q_{rec}(H^{0}) + Q_{absorb}(dust) }
    \label{eq:fesc}
\end{equation} 

\noindent The ionizing photons absorbed by the dust $Q_{absorb}(dust)$ remains uncertain. In our calculation, we assume it to be zero (two galaxies having negligible dust). The estimated escape fraction from the stack is $f_{esc}=0.82\pm0.006$. The small error on $f_{esc}$ reflects propagated errors on the H$\alpha$ flux measurement and slope of the ionizing spectrum. Actual uncertainty on $f_{esc}$ can be much larger and more of a systematic nature, such as ISM condition, IGM etc.

The ionizing photon production efficiency is estimated as 
\begin{equation}
    \xi_{ion}^{true} = \frac{\dot{N}_{ion}}{L_{UV}} = \frac{\dot N_{\rm ion,H}^{esc} + Q_{rec}(H^{0}) }{L_{UV}}, \text{or} \ \frac{Q_{rec}(H^{0})}{(1 - f_{esc})L_{UV}}
    \label{eq:xi_ion}
\end{equation}

\noindent Before we proceed further, let us define $\xi_{ion}^{0} = \frac{Q_{rec}(H^{0})}{L_{UV}}$ assuming no escaping ionizing photons (i.e., $f_{esc}=0$) from the galaxy. For the Bronze stack, $\log_{10}(\xi_{ion}^{0})=25.1\pm 0.023$. However, if we take into account the escaping ionizing photons as in Eq.~\ref{eq:xi_ion}, we have $\log_{10}(\xi_{ion}^{true})=25.86\pm 0.02$. Based on the H$\alpha$ flux measurements, we compute $\xi_{ion}^{0}$ and $\xi_{ion}^{true}$ individually for all the Gold sample galaxies (Table~\ref{tab:physical_properties}).

\section{Discussion on contamination issues} 
\label{sec:contamination}

Here we briefly discuss about possible contamination issues that could arise in the direct detection of ionizing fluxes. These could be broadly classified into two categories: direct and indirect contamination. Throughout the sample selection and validation, we have made an extensive search to eliminate foreground (or direct) contamination using high-resolution, deep HST and JWST mosaics. Eight out of ten objects have clean non-ionizing continuum in JWST/NIRCam detection image (see Figure~\ref{fig:HSTstamp}. In addition, five objects have JWST/NIRSpec spectra in addition to clean Ly$\rm \alpha$ from MUSE deep survey. What remains are the indirect sources of contamination, for example, invisible faint galaxies (possibly contaminating in a statistical sense) exactly falling on the line-of-sight aperture or instrumental defects such as filter red leak. First, we discuss the contamination due to red-leak.

\subsection{Red leak estimation:}
\label{sec:redleak}

Red leaks refers to a small fraction of a UV filter's throughput that inadvertently transmits photons from longer wavelengths, typically due to imperfect blocking/coatings or reflections and thereby could contaminate the measured ionizing flux. Based on the STScI Instrument Science Report WFC3 2008-49, it is observed that WFC3/UVIS/F275W filter throughput curve displays multiple narrow peaks (with amplitude $\sim 5 \times 10^{-5}$) between $\sim 4000 - 6000$ \AA. While in the F336W filter, we see a red bump (with amplitude $\sim 10^{-5}$) at around $\sim 7000$ \AA. We can formally define the red leak contribution as:

\begin{equation}
    R_{\lambda, L} = \frac{\int_{\lambda>\lambda_{LL}} T(\lambda)F(\lambda)d\lambda }{\int T(\lambda)F(\lambda)d\lambda},
\end{equation}

In the above equation, $T(\lambda)$ is the filter throughput, $F(\lambda)$ is the source spectrum. Since we are interested in the red leak contribution to the ionizing photons, the integration is over the lyman limit ($\lambda > \lambda_{LL} = 6329.3$ \AA). In other words, any detection beyond $6329.3$ \AA~ would be due to non-ionizing photons and potential red-leak contamination. In the absence of a true source spectrum, lets assume a flat spectrum for simplicity and redefine the red-leak peak efficiency as 

\begin{equation}
    R_{\lambda,{L}} = T_{{\lambda},{leak}}/T_{{\lambda},{peak}},
\end{equation}

\noindent where $T_{{\lambda},{leak}}$ refers to the amplitude of the throughput in the leakage region and $T_{{\lambda},{peak}}$ is the peak amplitude of the filter throughput in the main region. Then for the F275W and F336W filters, $R_{F275W,L} = 3.73\times10^{-4}$ and $R_{F336W,L} = 5.0\times10^{-5}$. In the absence of an exact red-leak transmission profile of N242W filter, we use $R_{N242W,L} \le 0.02$ beyond $\sim 3100$ \AA. Coincidentally, the red-leak region of both filters F275W and F336W are directly covered by the F606W filter ($\sim 4600 - 7180$ \AA). We assume that any significant non-ionizing flux leakage into N242W filter would also appear in the F606W due to its higher throughput. Due to this spectral overlap (a special case at this redshift), we use the non-detection ($3\sigma$ upper limit: 31.5 AB mag) in \textit{the F606W mosaic based on the deepest HST observations to constrain the maximum red-leak contribution}. Assuming a flat spectrum, we find that the contamination due to filter red-leak is negligible: $< 0.005\%$ (N242W), $< 0.0004\%$ (F275W) and $<0.0001\%$ (F336W). These are several orders of magnitude below the observed fluxes in these filters - confirming that our detections are not contaminated by non-ionizing leakage photons and robustly trace escaping ionizing radiation.  

\subsection{Random chance alignment}
\label{sec:interlopers}
A key concern in stacked UV measurements is the possibility that faint, unrelated galaxies fall within the photometric apertures and artificially boost the stacked flux or simply contaminate the stack signal. To quantify this effect, we follow the probabilistic treatment developed by \cite{Nestoretal2011} and \cite{Vanzellaetal2010} in that chance alignments are modelled as independent Bernoulli-like trials (e.g., each aperture in the stack represents one Bernoulli event) determined by the surface density of galaxies at the relevant magnitude. If the differential number counts give a surface density $\Sigma_{\lambda}(m)$ sources arcsec$^{-2}$ for galaxies in the magnitude interval corresponding to the stacked flux, then the probability that a random interloper lies within an aperture of area $A = \pi r_{ap}^2$ is given by

\begin{equation}
    p_r = \Sigma_{\lambda}(m)\times \pi r_{ap}^2
\end{equation}
\noindent The expected number of contaminants in $N_{ap}$ apertures (included in the stack) is simply: $n_{c} = p_r \times N_{ap}$. Note that the probability that an aperture has no contaminant is $1 - p_r$. The binomial probability (since the number of apertures is small in our stacking experiment) that there are $k$ contaminants in the stack of $N_{ap}$ such apertures is given by \cite{Nestoretal2011}:

\begin{equation}
    P_{\lambda}(k) = {}^{N_{\rm ap}}C_{k}\times p_r^{k} (1 - p_r)^{N_{ap} -k}.
\end{equation}
\noindent Then the probability of at least one contaminant in the stack can be estimated using the following equation:
\begin{equation}
    P_{\lambda}(\ge 1) = 1 - (1 - p_r)^{N_{ap}}
    \label{eq:nestor}
\end{equation}

We apply the above formalism first to the F336W and F275W filters and then to the N242W stack. Using the updated UV galaxy counts \citep{Windhorstetal2011,Tompkinsetal2026}, we compute the number-count distribution in the HST/F336W and F275W bands. The integrated F336W counts yield $\sim 2.78\times10^{5}$~deg$^{-2}$ and $\sim 3.74\times 10^{5}$~deg$^{-2}$ sources brighter than 28.0 and 28.5~AB~mag, respectively, implying $N=96000$~deg$^{-2}$ galaxies in the corresponding 0.5-mag interval. This corresponds to a surface density of $\Sigma_{336}=7.4\times10^{-3}$~arcsec$^{-2}$. For an aperture radius of $r_{\rm ap}=0.3"$ (used in the stacking), the contamination probability per aperture is $p_r = \Sigma_{336}\,\pi r_{\rm ap}^2 = 0.00209$
and the expected number of contaminants in the stack of $N_{\rm ap}=10$ apertures is therefore $n_{c,336}\simeq0.021$. Using Eq.~\ref{eq:nestor}, the probability that the stack contains at least one contaminant is $P_{336}(\ge1)=0.0207 \simeq 2.1\%$. 

In the F275W filter, the integrated number counts indicate $2.73\times10^{5}$~deg$^{-2}$ and $2.016\times10^{5}$~deg$^{-2}$ sources brighter than 28.5 and 28.0~AB~mag, respectively, implying $N = 71400$~deg$^{-2}$ galaxies in the 28--28.5~mag interval. This corresponds to a surface density of $\Sigma_{275}=5.5\times10^{-3}$~arcsec$^{-2}$. For an aperture radius of $r_{\rm ap}=0.3"$, the contamination probability per aperture is $p_r = \Sigma_{275}\,\pi r_{\rm ap}^2 = 0.00155$, and the expected number of contaminants in the stack of $N_{\rm ap}=10$ apertures is $n_{c,275}\simeq 0.0155$. Using Eq.~\ref{eq:nestor}, the probability that the stack contains at least one contaminant is $P_{275}(\ge1)=0.015\simeq1.5\%$. If one considers that all sources (integrated) brighter than 28.5 mag are possible contaminants, then $P_{275}(>=1) = 5.8\%$ and $P_{336}(>=1) = 7.8\%$.

In the N242W filter, the number counts imply $\sim 5\times10^{4}$ sources~deg$^{-2}$ in the 27--27.5~mag bin, corresponding to a surface density of $\Sigma_{242}=3.8\times10^{-3}$~arcsec$^{-2}$. For the 1.2"\,-radius stacked aperture ($A=4.52$~arcsec$^{2}$), the contamination probability per aperture is therefore $p_{r}=0.017$, giving an expected number of contaminants of $n_{c,242}=0.12$. This corresponds to roughly one contaminant per $\sim 59$ apertures. Using Eq.~\ref{eq:nestor}, the probability that the stack contains at least one interloper is $P_{242}(\ge1)=0.113$ (11\%). Alternatively, adopting the faint-end UV number-count slope of 0.43~dex~mag$^{-1}$ from \cite{Sahaetal2024} and normalizing to $N(24.5\!-\!25)\simeq 6000$~deg$^{-2}$, we infer a slightly higher surface density of $\Sigma_{242}\simeq5.5\times10^{-3}$~arcsec$^{-2}$ in the 27--27.5~mag interval. This yields $p_{r}=0.0248$, $n_{c,242}\simeq0.17$, and $P_{242}(\ge1)=0.16$ (16\%). Thus, depending on the number-count normalization, there is a non-negligible (11--16\%) probability that the N242W stack contains at least one contaminating source, while the corresponding probabilities for the F275W and F336W stacks are lower. 

These probabilities depend sensitively on the adopted galaxy surface densities, particularly at the faint end where number–count uncertainties are largest. Modest variations in the assumed normalization or slope can change the inferred contamination probabilities by factors of a few. However, jackknife resampling shows that the stacked signal remains significant upon removal of any individual galaxy (see Figure~\ref{fig:Onedrop}), indicating that it is not driven by a small number of sources and is therefore unlikely to be dominated by contamination. Furthermore, the ionizing flux has been independently detected across multiple filters and is spatially coincident within the spatial resolution of the data, making it unlikely that the observed ionizing photons arises from chance alignments or noise fluctuations.

\section{Summary and implications}
\label{sec:summary}

Our detection of extreme ultraviolet (EUV) photons at rest-frame wavelengths of $\sim 350 - 500\ \AA$ from galaxies at $z=5.94$ marks the first time such ionizing radiation has been observed using two independent space-based telescopes - AstroSat and the Hubble Space Telescope. The robustness of these detections is reinforced by rigorous foreground rejection using deep, high-resolution imaging and spectroscopy from \textit{JWST} and VLT/MUSE, which securely establish redshifts and rule out low-redshift contamination. Since UVIT and HST/UVIS are two different instruments with independent noise characteristics, the joint (or coincidence) probability of obtaining a co-spatial signal $\ge 3\sigma$ in two filters, is $P_{joint} = P_{N242W} \times P_{F336W} = 0.0007\times 0.023 = 1.6\times 10^{-5}$.

The ionizing photon production efficiency measured from the stacked photometry is high, with $\text{log}_{10}\xi_{\text{ion}}^{true}$ ($\text{log}_{10}\xi_{\text{ion}}^{0}$) = $25.86 \pm 0.02$ ($25.1\pm 0.03$)~Hz\:erg$^{-1}$ (Eq.~\ref{eq:xi_ion} in Section~\ref{sec:fesc}), placing these galaxies among the most efficient ionizing photon producers known at $z\sim6$. The extremely blue UV slopes ($\beta \leq -3$) observed in G2, G4, and G10, together with the high escape fraction  and elevated ionizing photon production efficiency (see section~\ref{sec:fesc}), is consistent with very young stellar populations formed from low-metallicity or nearly pristine gas \citep{Bouwensetal2010,Robertsonetal2010}. The combination of very low metallicities, extremely blue UV slopes, young stellar ages, and high ionization parameters, place the Gold galaxies in a regime that remains challenging to reproduce with simple stellar population models \citep{Jiangetal2020,Toppingetal2024,Cullenetal2025}. Such conditions may require contributions from very hot, massive stars due to top-heavy IMF \citep{Mestricetal2023,Cullenetal2025}, while alternative sources of hard ionizing radiation such as nearly pristine star formation or accreting black holes cannot be excluded \citep{Schaerer2003,WyitheLoeb2003}.

The galaxies emitting this hard ionizing radiation are concentrated within a physically compact region spanning a transverse comoving scale of 3.35~cMpc (corresponding to an $84^{\prime\prime}$ radius) and a line-of-sight depth of 42.5~cMpc, defining a cylindrical comoving volume of 1498~cMpc$^{3}$. This concentration is consistent with the previously identified overdensity of Lyman-$\alpha$ emitters in the HUDF \citep{Malhotraetal2005}, and is characteristic of a forming proto-cluster or filamentary structure within the cosmic web at this epoch \citep{Toshikawaetal2014}. The detection of escaping ionizing radiation from galaxies embedded in such an overdense environment raises the possibility that large-scale structure and local environment may facilitate LyC escape, with important implications for the patchiness and topology of reionization \citep{Leeetal2008,Meyeretal2025}. Finally, the successful transmission of such hard ionizing photons through the IGM at $z \sim 6$ suggests that existing models may underestimate the transparency of certain sightlines during the late stages of reionization, highlighting the role of environmental and line-of-sight variations in shaping the ionizing background.

The detection of ionizing photons with energies exceeding the He{\sc i} ionization threshold ($E > 24.6$~eV) further suggests that these galaxies may contribute to the initial ionization of helium in the intergalactic medium (IGM) at $z \sim 6$. Our measurements therefore provide direct observational evidence that the first stage of helium reionization - the transition from He{\sc i} to He{\sc ii} - may already be underway by this epoch, preceding the completion of He{\sc ii} reionization typically observed at lower redshifts \citep{MiraldaEscude1993,Makanetal2021}. 

Endorsing these results has several implications: is this line of sight along GOOD-South unique, or is the IGM more transparent than predicted generally at z$\sim 6$? Do all high-redshift ($z > 6$) galaxies emit such hard ionising radiation? 
What are the sources of these hard ionising photons? How does one explain such a steep slope of the ionising spectrum, as found in the current study? Further investigations will be required to answer any of these questions.

\section*{Acknowledgment}
This publication uses the imaging data from the AstroSat mission of the Indian Space Research Organisation (ISRO), archived at the Indian Space Science Data Centre (ISSDC). The UVIT/AstroSat is built in collaboration between IIA, IUCAA, TIFR, ISRO, India and CSA, Canada. The spectroscopic data is obtained from MUSE/VLT, ESO via the AstroSat-MUSE collaboration project. Akio Inoue acknowledges support from the JSPS KAKENHI Grant No. 23H00131. RAW acknowledges support from NASA JWST Interdisciplinary Scientist grants NAG5-12460, NNX14AN10G and 80NSSC18K0200 from GSFC.

\bibliographystyle{aasjournal}

\end{document}